\documentclass[fleqn,usenatbib]{mnras}

\usepackage{newtxtext,newtxmath}

\usepackage[T1]{fontenc}

\DeclareRobustCommand{\VAN}[3]{#2}
\let\VANthebibliography\thebibliography
\def\thebibliography{\DeclareRobustCommand{\VAN}[3]{##3}\VANthebibliography}

\usepackage{graphicx}	
\usepackage{amsmath}	
\usepackage{longtable}
\usepackage{multirow}
\usepackage{supertabular,booktabs}
\usepackage{lscape}
\usepackage{rotate} 
\usepackage{hyperref}
\usepackage{float}

\usepackage{ulem}

\DeclareRobustCommand*{\N}{\textit{NuSTAR}}
\newcommand{\NS}{\textit{NuSTAR\_Spectra}}

\newcommand{\nup}{$\nu_{\rm peak}$}
\newcommand{\nufnu}{$\nu$f$({\nu})$}

\title[The first \textit{NuSTAR} blazars catalogue]{The first hard X-ray spectral catalogue of Blazars observed by \textit{NuSTAR}}

\author[R. Middei et al.]{
Riccardo Middei,$^{1,2}$\thanks{E-mail: riccardo.middei@ssdc.asi.it}
Paolo Giommi,$^{3,4,5,6}$
Matteo Perri, $^{1,2}$
Sara Turriziani, $^{7}$
Narek Sahakyan, $^{8,6,9}$ 
\newauthor 
Y. L. Chang $^{10}$
C. Leto $^{1,11}$
F. Verrecchia $^{1,2}$
\\
$^{1}$ Space Science Data Center, SSDC, ASI, via del Politecnico snc, 00133 Roma, Italy\\
$^{2}$ INAF - Osservatorio Astronomico di Roma, via Frascati 33, I-00040 Monteporzio Catone, Italy\\
$^{3}$ Institute for Advanced Study, Technische Universit{\"a}t M{\"u}nchen, Lichtenbergstrasse 2a, D-85748 Garching bei M\"unchen, Germany\\
$^{4}$ Center for Astro, Particle and Planetary Physics (CAP3), New York University Abu Dhabi, PO Box 129188 Abu Dhabi, United Arab Emirates;\\
$^{5}$ Associated to Italian Space Agency, ASI, via del Politecnico snc, 00133 Roma, Italy\\
$^{6}$ ICRANet, P.zza della Repubblica 10, 65122, Pescara, Italy\\
$^{7}$ Physics Department, Gubkin Russian State University (National Research University), 65 Leninsky Prospekt, Moscow, 119991, Russian Federation\\
$^{8}$ICRANet-Armenia, Marshall Baghramian Avenue 24a, Yerevan 0019, Armenia\\
$^{9}$ICRA, Dipartimento di Fisica, Sapienza Universit\`a di Roma, P.le Aldo Moro 5, 00185 Rome, Italy\\
$^{10}$Tsung-Dao Lee Institute, Shanghai Jiao Tong University, 800 Dongchuan RD. Minhang District, Shanghai, China\\
$^{11}$ASI - Italian Space Agency, Via del Politecnico snc, 00133, Rome, Italy\\
}

\date{Accepted XXX. Received YYY; in original form ZZZ}

\pubyear{2021}

\begin{document}
\label{firstpage}
\pagerange{\pageref{firstpage}--\pageref{lastpage}}
\maketitle

\begin{abstract}

Blazars are a peculiar class of active galactic nuclei (AGNs) that enlighten the sky at all wavelengths. The electromagnetic emission of these sources is jet-dominated resulting in a spectral energy distribution (SED) that has a typical double-humped shape. X-ray photons provide a wealth of information on the physics of each source as in the X-ray band we can observe the tail of SED first peak, the rise of the second one or the transition between the two. \N, thanks to its capability of focusing X-rays up to 79 keV provides broadband data particularly suitable to compute SEDs in a still poorly explored part of the spectrum. In the context of the Open Universe initiative we developed a dedicated pipeline, \textit{\NS}, a shell-script that automatically downloads data from the archive, generates scientific products and carries out a complete spectral analysis. The script homogeneously extracts high level scientific products for both \N's telescopes and the spectral characterisation is performed testing two phenomenological models. The corresponding X-ray properties are derived from the data best-fit and the SEDs are also computed. The systematic processing of all blazar observations of the \N\ public archive allowed us to release the first hard X-ray spectroscopic catalogue of blazars (\textit{NuBlazar}). The catalogue, updated to September 30th, 2021, includes 253 observations of 126 distinct blazars, 30 of which have been multiply observed.
\end{abstract}




\section{Introduction}
\begin{figure}
\includegraphics[width=0.49\textwidth]{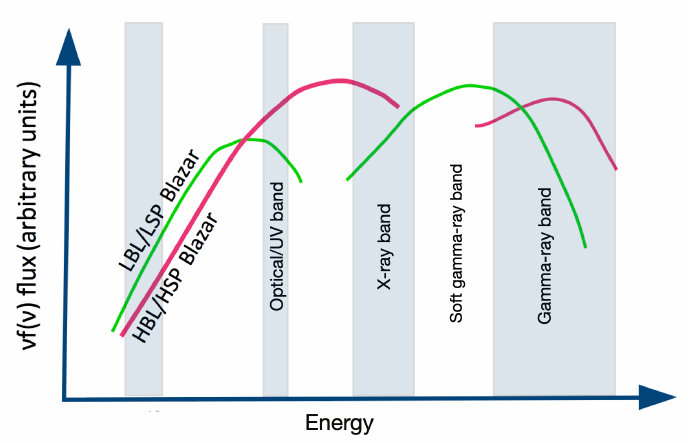}
\caption{\small{Blazar SEDs shown for different peak energies of the underlying two main emission bumps. LBLs sources have the maximum of their synchrotron power output in the infrared/optical band, while for HBL the peak occurs between the UV or X-ray domain.	Moreover, the SED of an extreme class of blazars, the so-called ultra high energy BL Lacs (UHBLs) for which the maximum energy output of the synchrotron emission is well above the hard X-rays domain. It is worth noticing that for the same peak luminosity, the radio emission decreases by orders of magnitudes. Figure adapted from \citet{Giommi2002babs}.\label{sed}}}
\end{figure}
Active Galactic Nuclei (AGNs) are powerful, persistent extra-galactic sources powered by accretion onto a supermassive black hole (SMBH, 10$^6$-10$^{10}$ M$_{\sun}$). Their broadband spectrum results from the interplay of thermal and non-thermal mechanisms that involve different structures from sub-parsec  up to kilo parsecs scales \citep[e.g.][]{Bianchi2012,Netzer2015,Padovani17}.\\
\indent Blazars are a rare and highly peculiar class of AGN as they display a narrow relativistic jet that happens to be oriented towards the Earth. They radiate from radio wavelengths to the $\gamma$-ray domain \citep[e.g.][]{Urry95}, and, possibly, they also emit high-energy neutrinos \citep[][]{neutrino,Giommi2020,GiommiPadovani2021}. 
These sources are radio bright, show prominent flux variability and sometimes large polarisation signals and they can be classified via physical and geometrical conditions as discussed by \cite{Giommi2012}. The optical spectral properties of these sources are used to define two major classes, Flat-Spectrum Radio Quasars (FSRQs) and BL Lacertae objects (BL Lacs). Objects belonging to the first group, similarly to radio quiet AGNs, show broad emission lines, while a strongly variable and very often featureless continuum is typical of BL Lacs.\\
\indent The spectral energy distribution (SED) of blazars has a characteristic double-peaked shape, see Fig.~\ref{sed}. The low energy part, normally assumed to be due to synchrotron emission, peaks at a frequency (\nup\,) that is located between far infrared and X-rays. This frequency is used to further classify blazars into low, intermediate, and high energy peaked, LBL (\nup\, < 10$^{14}$ Hz), IBL (\nup\, is between 10$^{14}$ and 10$^{15}$ Hz), and HBL (\nup\,>10$^{15}$ Hz), respectively \citep{Padovani1995,Abdo2010}. 
The physical mechanism that generates the high energy part (commonly observed between the X-rays and the $\gamma$-rays) is still a debated matter. In the leptonic interpretation, two competing scenarios have been proposed, i.e., inverse Compton (IC) scattering of photons provided by the synchrotron emission \citep[Synchrotron Self Compton, SSC; e.g.][]{Maraschi1992,Mastichiadis2002} or produced outside the jet \citep[ e.g.][]{Dermer93,Sikora1994,Blazejowski2000}, with this often referred as External Compton (EC). The most likely external photon fields can be the photons reflected from the broad-line region or photons from a dusty torus. In the alternative hadronic scenario, the high energy peak is either due to proton synchrotron radiation \citep{Aharonian2000} or to emission processes of secondary particles produced in hadronic interactions in the jet \citep{Mannheim1993,Mucke2003}.
Blazars have a quite unique behaviour compared to other types of AGNs \citep[see e.g.][]{Maccacaro1984,Sedentary,Rector2000,ST2019blaz}, as showed by the analysis of samples of X-ray selected blazars where no or even negative cosmological evolution were found \citep[][]{Rec00,Wol01b}, and the same trend was discussed in \cite{ajello2014} for a sample of $\gamma$-ray selected sources.\\
\indent In this paper we consider all blazars observed by the high energy satellite \textit{NuSTAR}. Hard X-rays carry important information about blazars as shown by earlier \textit{BeppoSAX}-based analyses \cite[e.g.][]{Tavecchio2000,Massaro2004,Massaro2004a,Donato2005}, and by works based on INTEGRAL \cite[e.g.][]{Gianni2011} and Suzaku data \cite[e.g.][]{Lukasz2014,Zhang2019,Zhang2021}, and numerous papers rely on \N~data \cite[][]{Sbarrato13,Sbarrato16,Marcotulli17,Bhatta2018,Ghisellini19}. However, an extensive study of the hard X-ray properties of blazars observed by \N~has never been performed since most of the characterisations focuses on single sources or small samples. In an effort to facilitate sample studies we performed a systematic analysis of all the blazars included in the public \N\,archive, reporting their spectral properties in the first 
\N\,blazars catalogue (\textit{NuBlazar}). Such a catalogue is based on the \NS\, script, a software tool similar to Swift\_xrtproc \citep{GiommiXRTspectra}, both developed in the context of the Open Universe (OU) initiative \citep{GiommiOU} in collaboration with Space Science Data Center (SSDC) of the Italian Space Agency (ASI).\\

\section{Blazars in the \textit{NuSTAR} public archive}

\begin{figure}
	\includegraphics[width=0.49\textwidth]{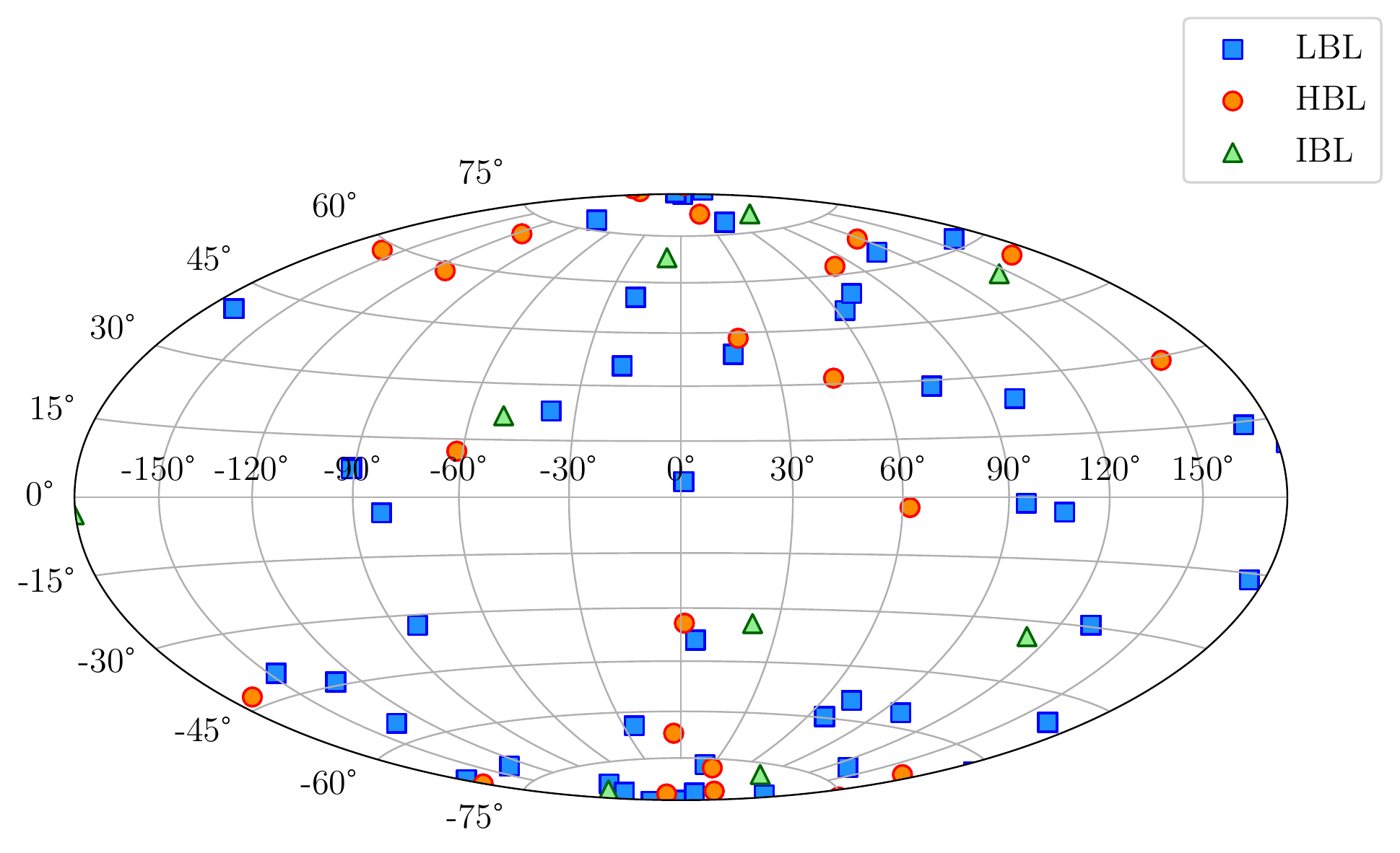}
	\caption{\small{Hammer projection in galactic coordinates of all blazars observed with \N and studied in this work. For the sake of simplicity, we plotted uncertain type blazars in green as IBL sources.}}\label{hammer}
\end{figure}

\textit{NuSTAR} \cite[][]{Harr13} is a NASA Small Explorer (SMEX) mission consisting of two co-aligned grazing incident hard-X-ray telescopes that collect light onto the two focal plane solid state detectors (FPMA/B). Thanks to its $\sim$10 metres focal length mirror-detector design, \N\ is the first focusing high-energy X-ray mission, hence the first facility for which hard X-ray spectra are not background dominated.
The \N\ broadband-pass is extremely suitable for addressing a number of different issues in AGNs physics, including the characterisation of the X-ray emission in blazars. In these sources, in fact, different components can be observed in the operating 3-79 keV energy range: the high frequency tail of the synchrotron peak in HBL blazars, the rising SSC component in LBL objects, or the transition between the low and high energy components. In the case of FSRQs dominated by EC emission \citep[see e.g.][]{Bo2013}, \N\ data can be used to put constraints on the EC component, if the transition between SSC and EC components happens in the hard X-ray band \citep{Sikora2013}.\\
\indent To compile the list of blazars included in the \N~public archive, we cross-matched the Open Universe master list of blazars and blazar candidates \citep{Giommi2019} with the list of all the \N\ observations carried out before the end of September 2021. The master list combines the \textit{BZCAT} 5th edition \citep{Massaro2015,Massaro2016Cat}, the sample of blazars detected by Fermi LAT and reported in the 4LAC-DR2 catalogue \citep{Ackermann2015,Ackermann2015Cat} and the 3HSP high energy peaked and extreme blazars list \citep{Chang2019,Chang2019Cat}.\\
Our initial sample of \N\, observations of blazars consisted of 287 pointings including each of the sources in the master list within 7 arcmin from the center of the field of view. We then neglected all those observations with an exposure shorter than 2000 seconds in total for the two \textit{NuSTAR} modules and those fields in which the background level was different between the \textit{FPMA} and \textit{FPMB} modules.\\
\indent The remaining matches included sources such as NGC 1275, M87 and Centaurus A. These sources are located in the Perseus, Virgo and Centaurus clusters, respectively, thus their X-ray emission detect by \N\, is due to a nuclear and an extra-nuclear contribution emerging from the hot intracluster gas. The complex nature of these spectra is unsuitable for an automatic analysis, thus we did not consider observations belonging to any of these three sources. Finally, we excluded fields in which the source, if present, was characterised by a S/N too low to allow for a spectral characterisation.\\
\indent These steps resulted in a final list of 253 exposures of 126 blazars. In
Fig.~\ref{hammer} we show a sketch of the positions of the sources of the sample, whose details are given in Table~\ref{listsources}.

\section{Data processing: the \NS\, pipeline}

To reduce and analyse in a homogeneous and automatic way the selected observations we developed the \NS\, pipeline, a shell-script based on the \textit{NuSTAR} Data Analysis Software (NuSTARDAS\footnote{Developed at the ASI/SSDC}), included in the \textit{HEASoft} V6.29 package. The calibration data files distributed with CALDB version 20211020 were adopted. Using the \NS\, pipeline it is possible to download data, generate high-level calibrated scientific products and perform spectral analyses via a simple command line.\\
\indent As outcomes, \NS\, returns the FPMA/B source and background spectra, the corresponding ancillary and response files, the best-fit parameters for a power-law and a logarithmic parabola model, integrated fluxes in the 3-10 and 10-30 keV bands, and the best fit spectral data in \nufnu\, units for SED plotting.  Moreover, at the end of the procedure, various plots in gif format are created to facilitate visual inspection of the results of the analysis and of the products generated.\\

\begin{figure}
	\includegraphics[width=0.48\textwidth]{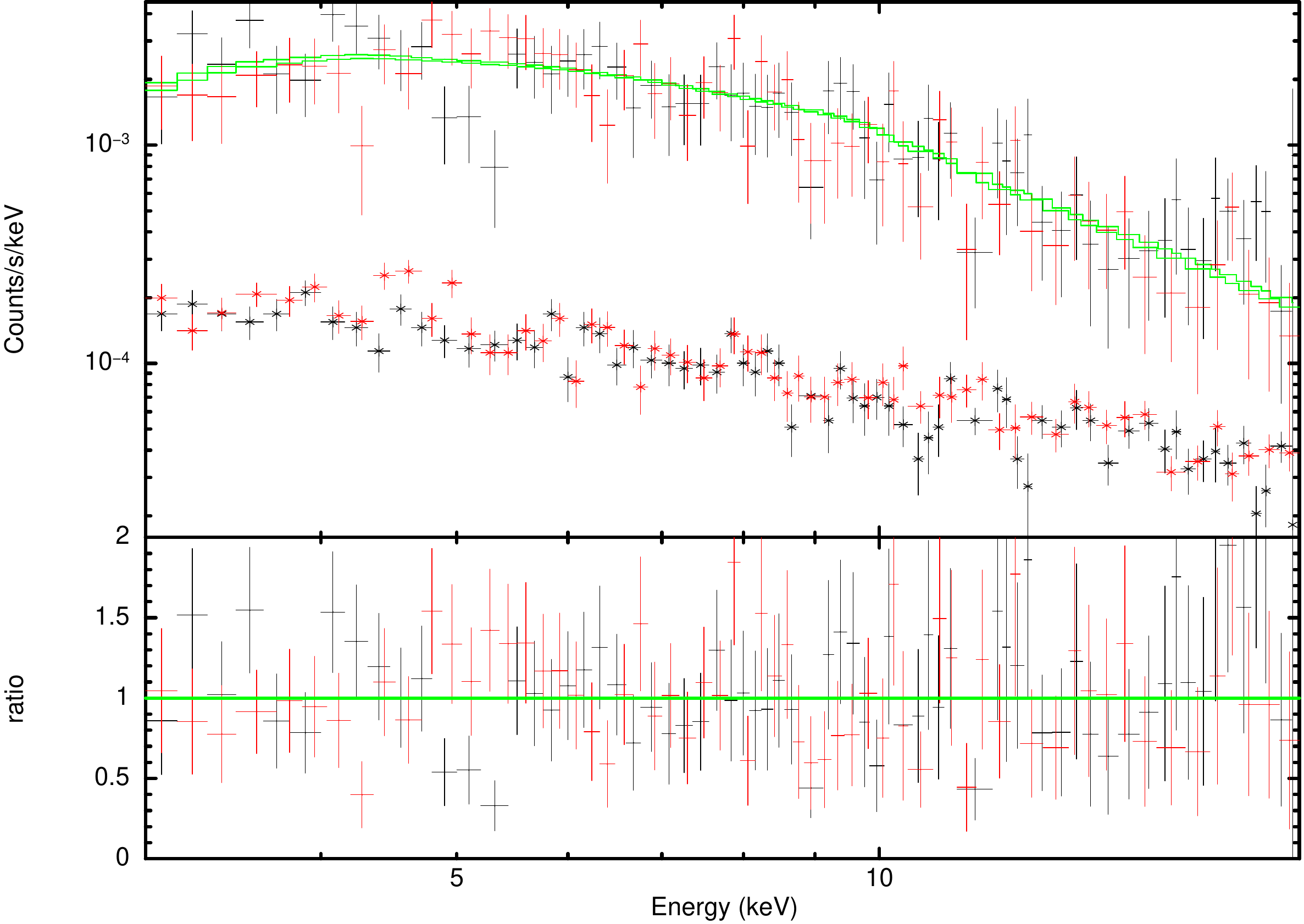}
	\includegraphics[width=0.49\textwidth]{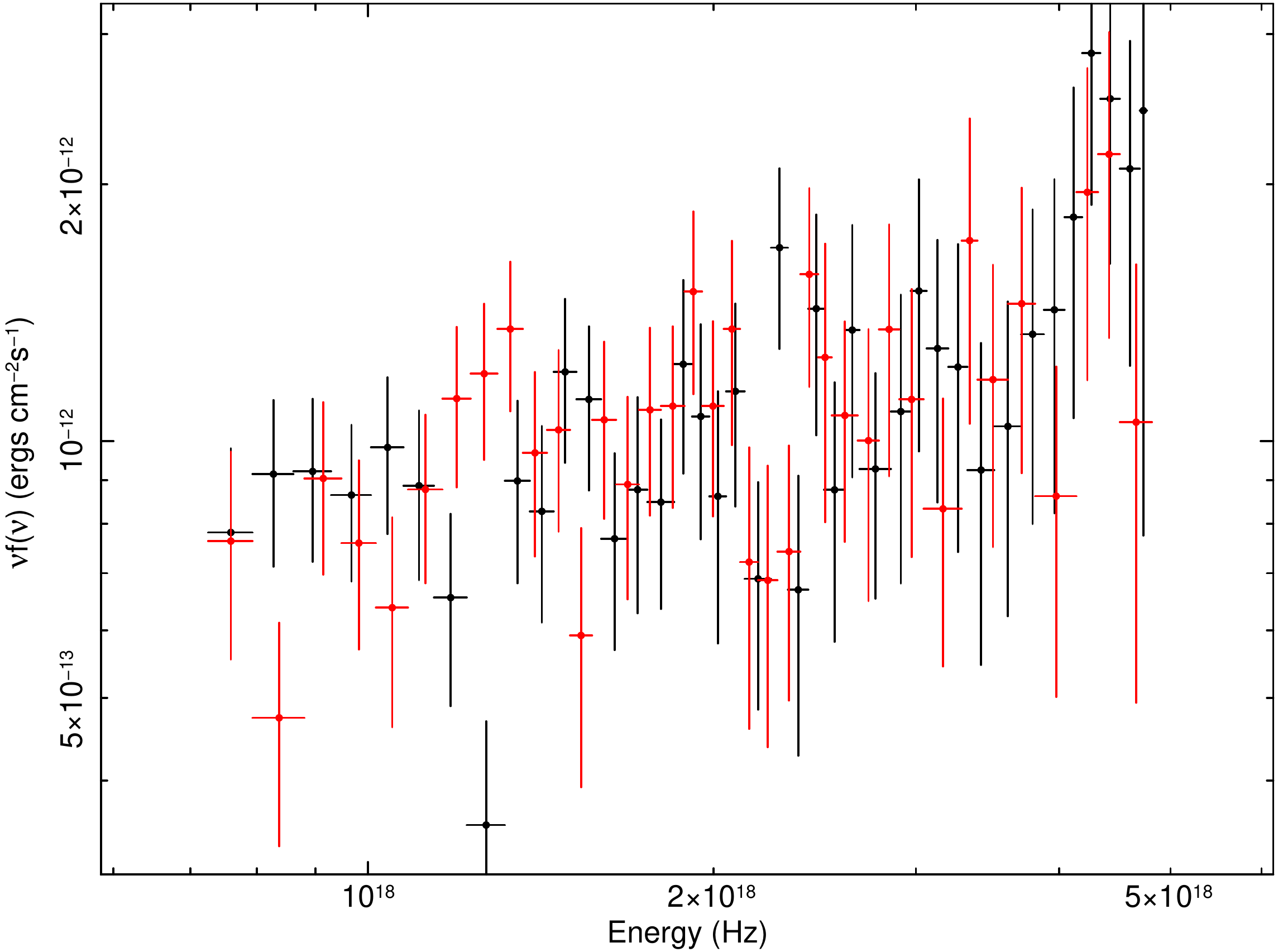}
	\caption{\small{Top panel: The spectral fit to the FPMA (black) and FPMB (red) spectra of the source 5BZB J0509+0541 (Obs. ID 090402637004). Background spectra are also showed. The model (green) accounts for a power-law absorbed by the Galaxy. In this case, the maximum energy for the analysis correspond to 20 keV. Bottom panel: the corresponding spectral energy distribution is showed. Similar plots can be found for all the sources studied in the present paper at the web page \url{https://www.ssdc.asi.it/nustarblaz/}.}}\label{showexample}
\end{figure}

\noindent The following steps are executed within the \NS~script:

\begin{itemize}
	\item calibrated and filtered event files are downloaded from the SSDC repository\footnote{ \url{ftp://archive.ssdc.asi.it/nustar/pubdata/}} and saved into a folder named as the observation ID. The script does not perform an ex-novo data calibration (level 1 data) and screening (level 2 data) but relies on the level 2 event files generated at the \textit{NuSTAR} SOC with default parameter settings and available for each observation ID directly from HEASARC (or SSDC mirror site) in the archive.
	\item the ftool \textit{fselect} \citep{Blackburn1995} is used to read the event file and generate a new file in the 3-20 keV energy range. In this interval, the effective area of the modules \textit{FPMA} and \textit{B} is maximal and the background contribution is minimised. Using this file, the source count rate is derived.
	\item the \textit{detect} task of the X-ray image package \textit{Ximage} is launched to  precisely locate the source in celestial coordinates. If no source is detected the script reports on this issue and stops.
	\item High-level science products for the detected source are extracted using the \textit{nuproducts} routine. The source counts are extracted using a circular region, while the background is computed in an annulus centered on the source. The extraction radius/radii for the circle/annulus are automatically set to a value that is optimised depending on the source count rate. The higher/lower the count rate the larger/smaller will be the radius of the circular region which usually varies in the range of 30-70 arcsec. In a similar fashion, the annular region also scales maintaining a minimum separation of 50 arcsec between the inner and outer radii. The ancillary and response matrices are computed at this stage and the spectra are grouped via the \textit{grppha} standard command in order to have at least 1 count for bin.
	\item A spectral analysis is performed using the \textit{Xspec} fitting package  \citep[][]{Arna96} adopting Cash statistics \citep{Cash79}. The fit is performed from 3 keV up to the maximum energy where the signal is still present, typically between  20 and 79 keV. Spectra of module A and B are simultaneously fitted with a cross-calibration constant \cite[found to be always within a few percent,][]{Madsen2015} and the Galactic column density is always taken into account with the \textit{tbabs} model. Then, the fitting model can be described as:
\textit{tbabs$\times$const$\times$model}, where the latter term stands for power-law or logarithmic parabola. The logarithmic parabola is defined as 
\begingroup
\begin{equation}
\rm N( E)=( E/ E_1)^{-\alpha-\beta\log( E/E_1)},
\label{eq}
\end{equation}
\endgroup
where $\alpha$ accounts for the source photon index at the energy E$_1$, $\beta$ represents the curvature of the parabola and E$_1$ is a reference energy.\\
When we process the data we obtain the best-fit values of normalisation and 
photon index for the power-law model, and the spectral index and the curvature for the logarithmic parabola. Finally, E$_1$ is set to 5 keV and kept fixed in agreement with \citet[][]{Massaro2004} and other \N\, studies \citep[e.g.][]{Balokovic2016}. 
Once the best parameters are estimated the script saves the observed flux in the 3-10 and 10-30 keV bands and the current best-fit in \textit{Xspec} format. 
The choice for the flux ranges was driven by the effective area of the modules \textit{FPMA} and \textit{B} detector that is maximal between 3-20 keV.

\item \NS\, uses the best-fit parameters to compute the X-ray SED from 3 keV up to the energy used to perform the spectral analysis. SEDs, are calculated separately for the two models. In particular, the power-law best-fitting the \textit{FPMA/B} spectra is used to derive the SED of each observation. Though the \N\ band is only marginally affected by absorption, we de-absorbed the fluxes using the cross sections of \citet{Morrison1983} and setting the amount of absorbing material (N$_{\rm H}$) equal to the Galactic value, the same value assumed for the fits.
\item Finally, the pipeline saves gifs and other files of interest, i.e. those allowing for a different spectral analysis, and those containing the information on the
flux and spectral parameters. In Fig. \ref{showexample}, we show an example of a spectral fit performed on 5BZB J0509+0541 (Obs. ID 090402637004) and the corresponding X-ray SED.
\end{itemize}
\section{The NuBlazar catalogue}
\begin{figure}
	\includegraphics[width=0.49\textwidth]{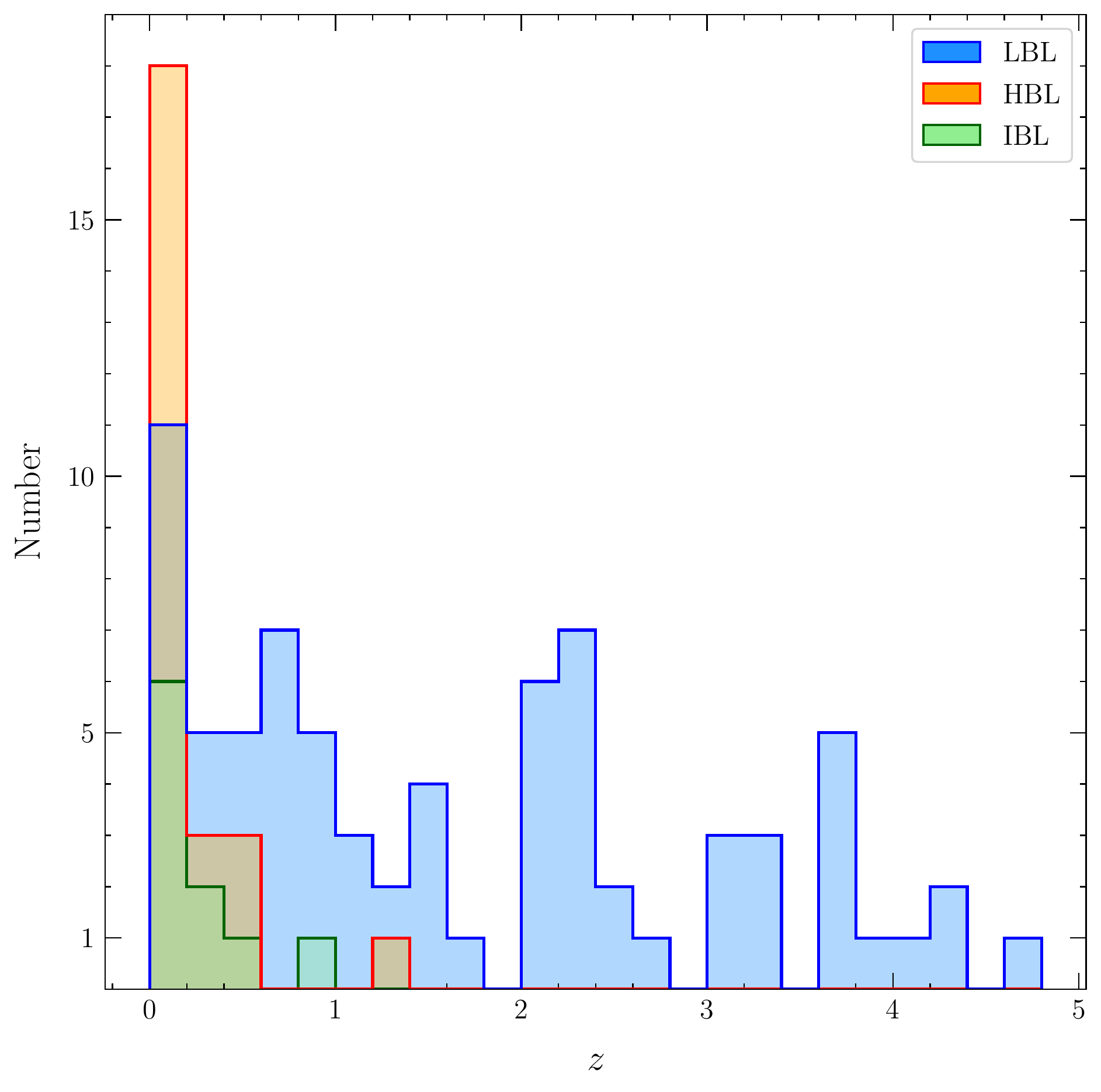}
	\caption{\small{Redshift distributions for the different classes of blazars listed in the NuBlazar.}}\label{isto}
\end{figure}
\begin{figure}
	\centering
	\includegraphics[width=0.49\textwidth]{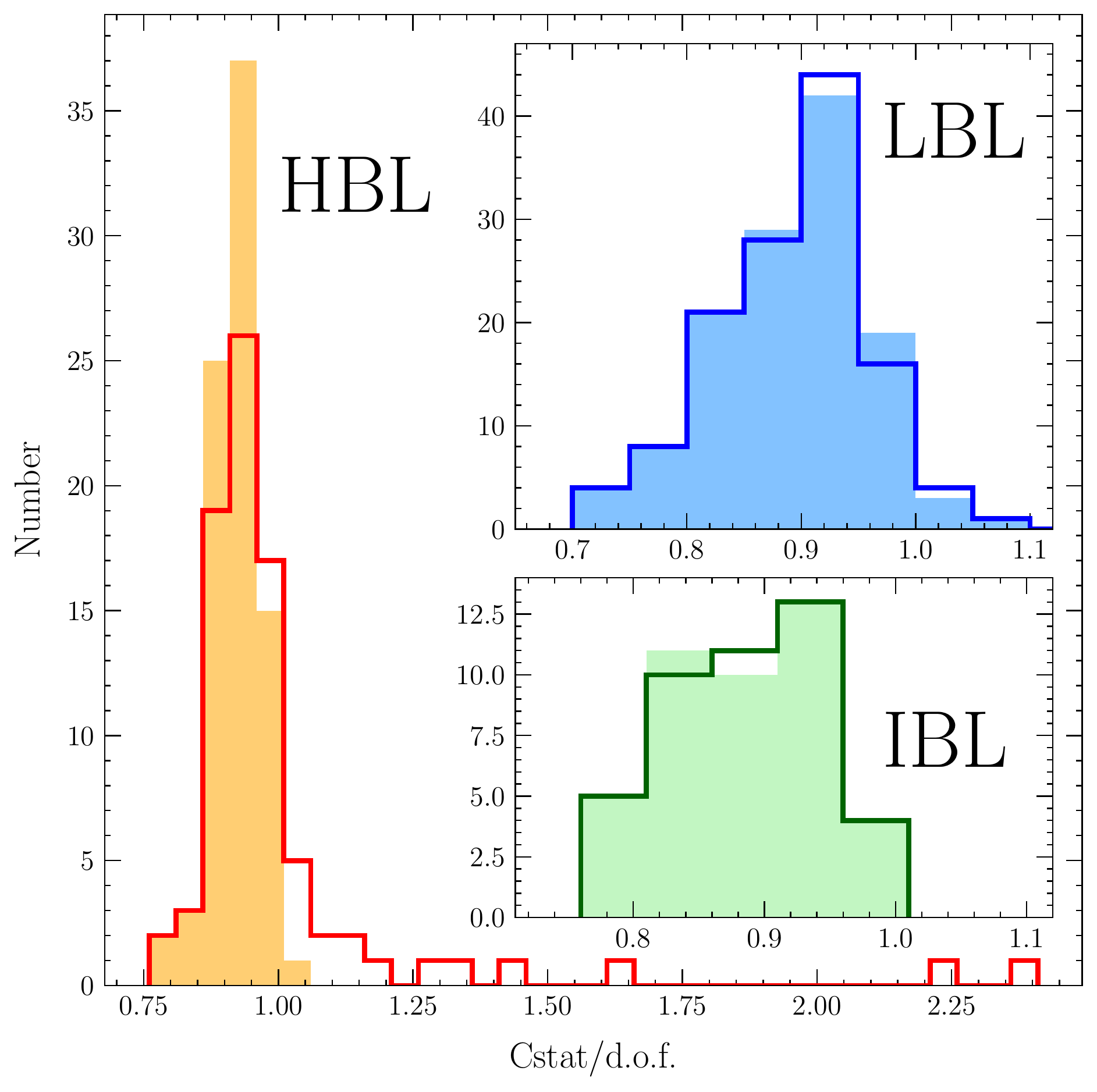}
	\caption{\small{Distributions of the best fit statistics for the different sub-classes of blazars in our catalogue. Solid lines indicate the distributions corresponding to the power law model, while filled histograms refer to the log parabola model.}\label{statistics}}
\end{figure}
\begin{figure}
	\includegraphics[width=0.49\textwidth]{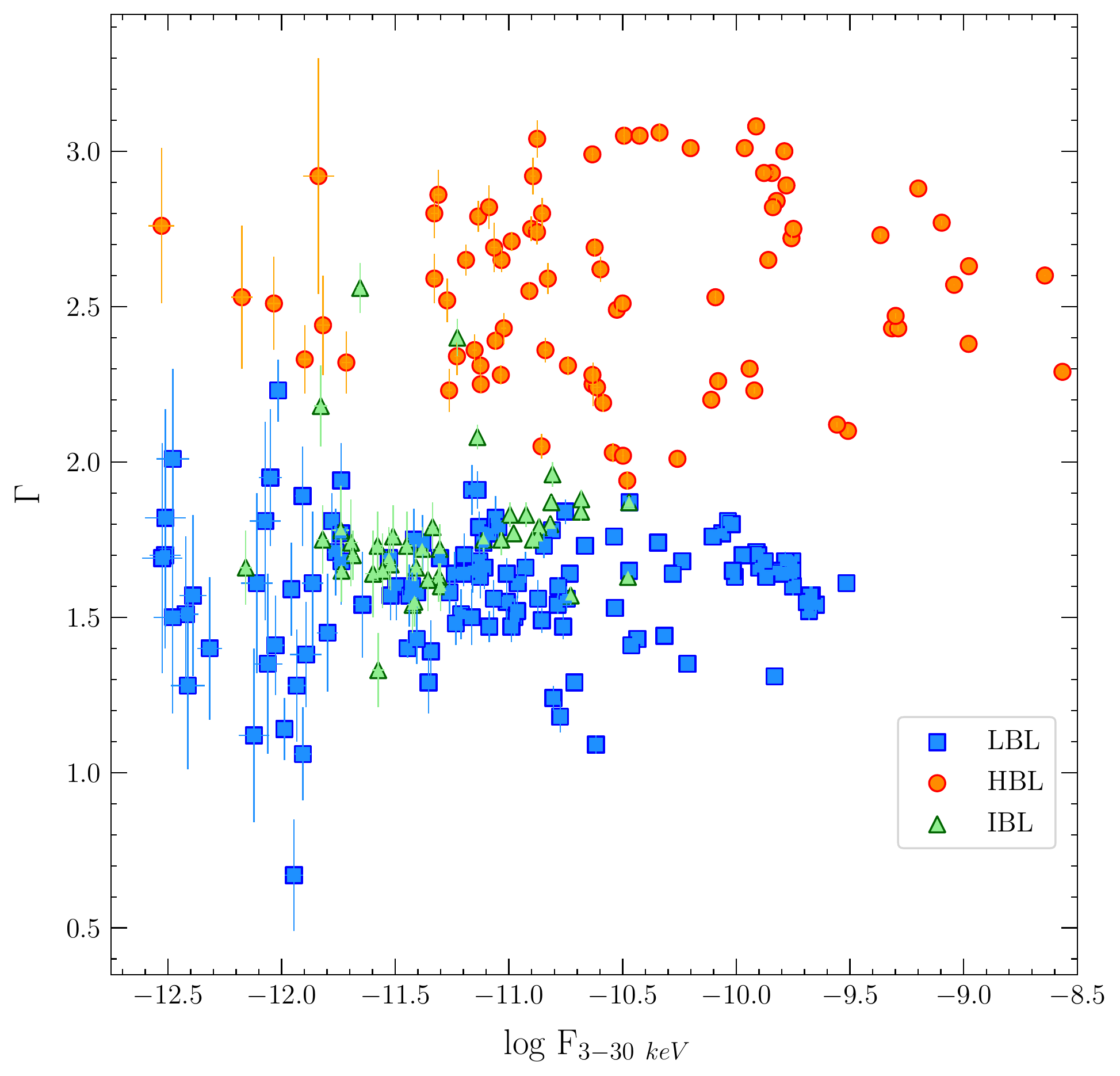}
	\caption{\small{Photon index distributions for the different classes of blazars. HBLs have steeper spectra than LBL sources. The photon indices of IBLs are mostly flat, with occasional steeper values}}\label{gammaflux}
\end{figure}
\begin{figure}
	\includegraphics[width=0.49\textwidth]{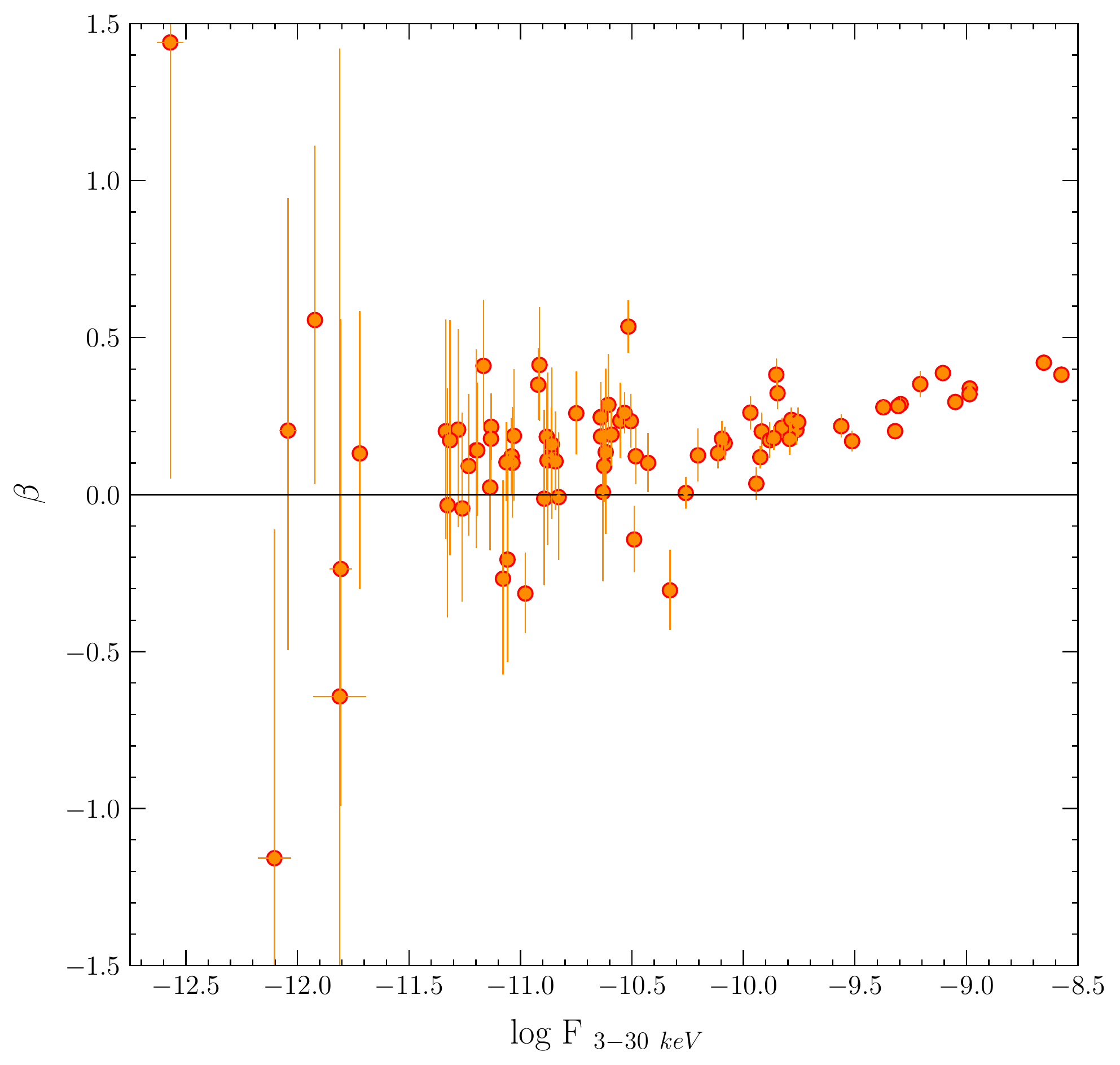}
\includegraphics[width=0.49\textwidth]{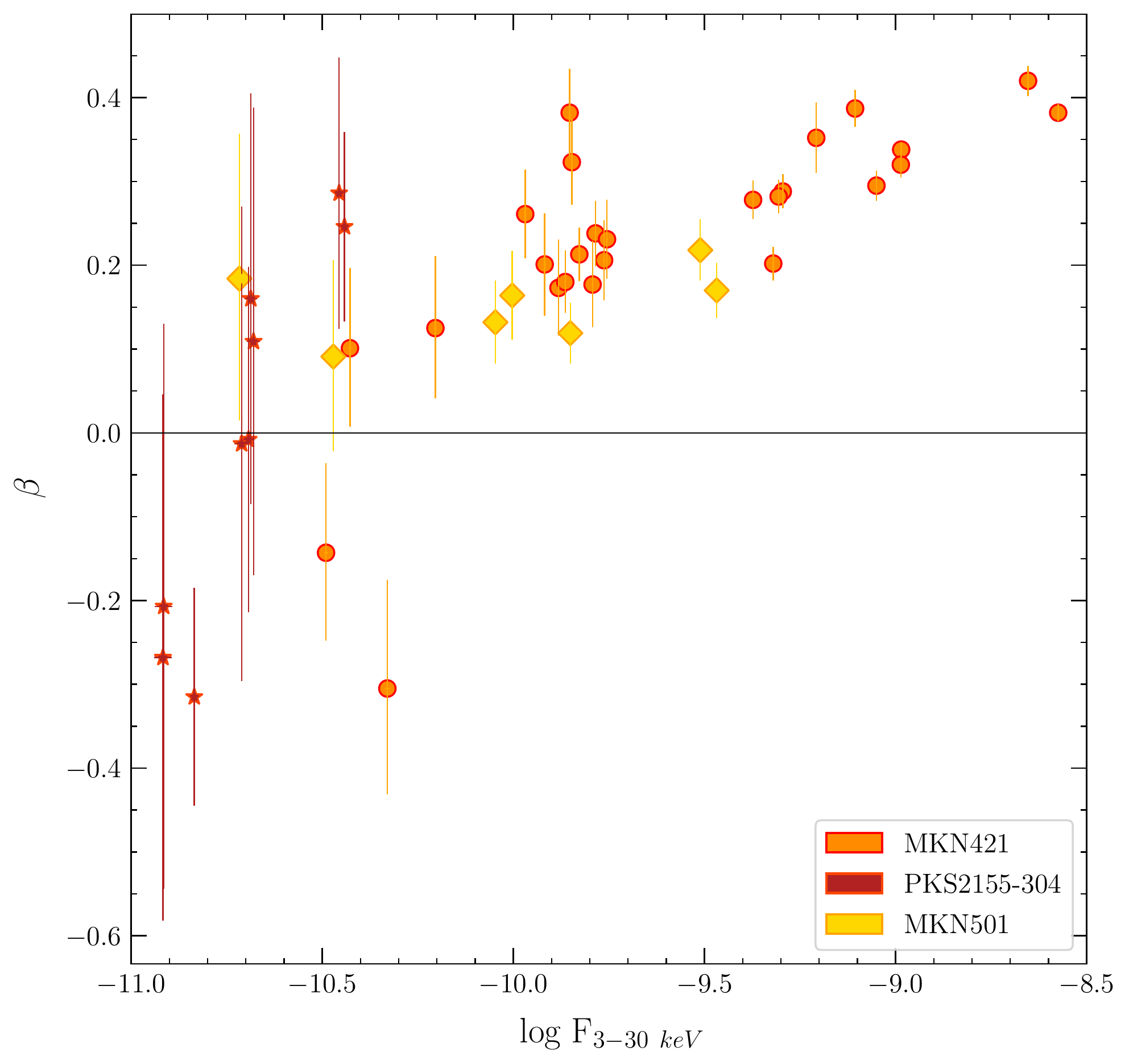}
	\caption{\small{\textit{Top panel}: curvature parameter is plotted against flux for HBL blazars. \textit{Bottom panel}: Curvature parameter for the three sources labelled in the legend. The $\beta$ parameter is positively correlated with the source flux in Mkn 421 and PKS 2155+304 while in Mkn 501 this trend is not observed.}}\label{beta}.
\end{figure}
\begin{figure}
 \includegraphics[width=0.49\textwidth]{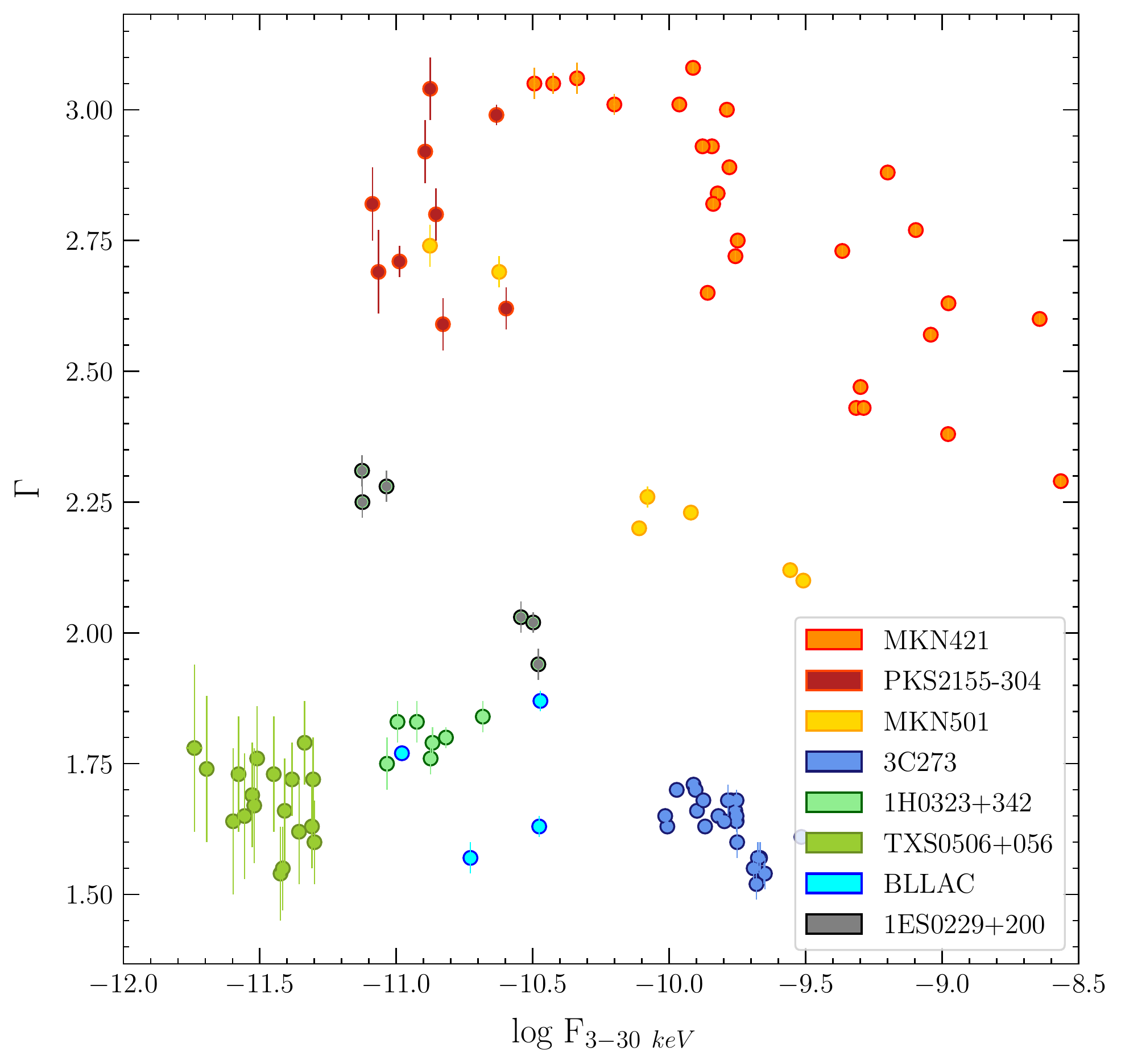}
	\caption{\small{The photon index from the power law of the labelled sources is plotted as a function of the observed flux in the 3-30 keV band. For instance, both Mkn 421 and Mkn 501 show a typical harder when brighter behaviour.}}\label{gamma1}.
\end{figure}
\begin{figure}
	\includegraphics[width=0.49\textwidth]{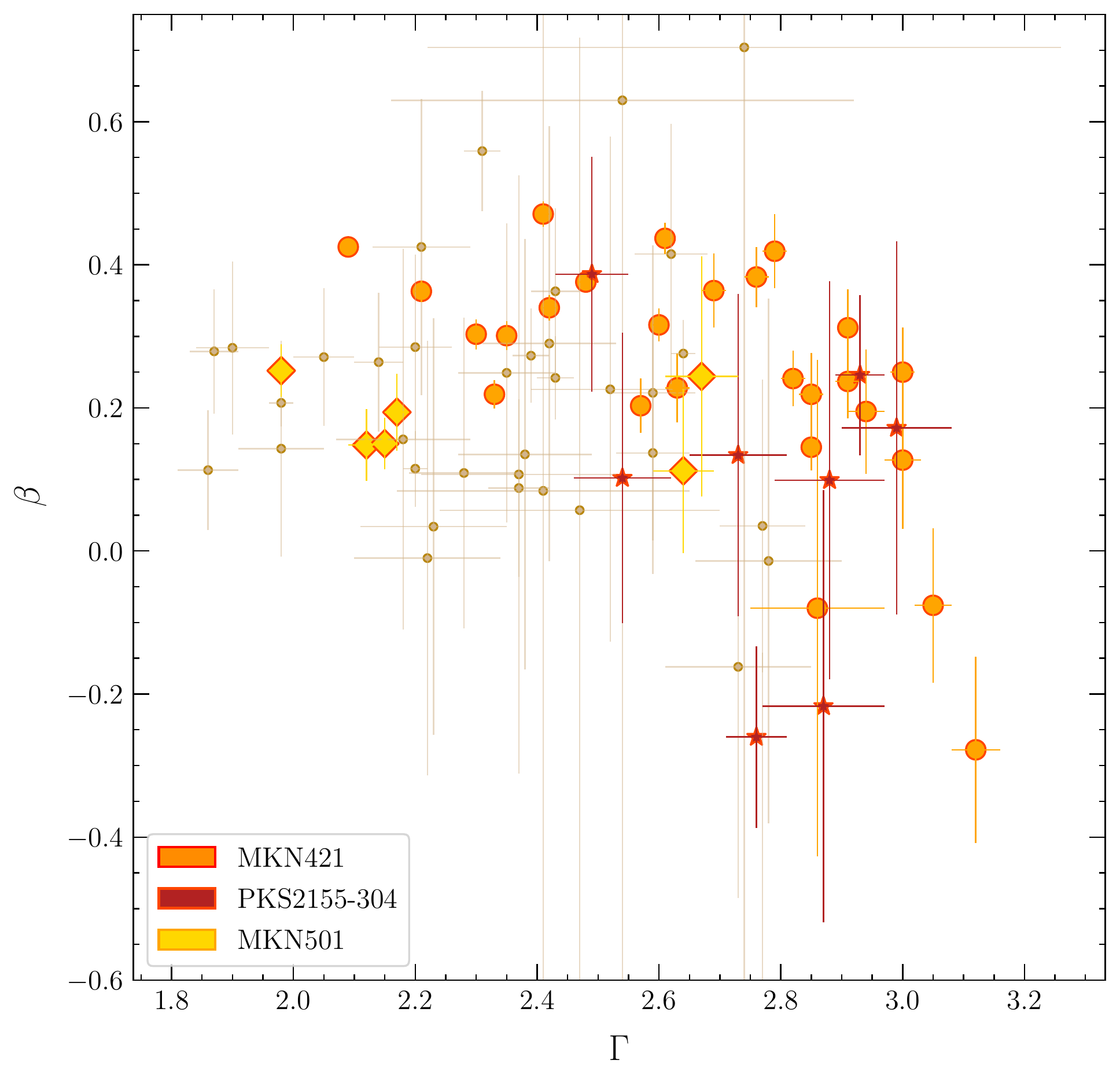}
	\caption{\small{Spectral curvature as a function of the primary photon index for the HBL sources.}}\label{isto2}
\end{figure}

The catalogue resulting from our processing is called \textit{NuBlazar} and presents the spectral and observational properties of 126 individual blazars, 30 of which have been multiply observed, for a total of 253 exposures. \textit{NuBlazar} includes 80 LBLs, 32 HBLs and 14 blazars with intermediate SEDs (IBL). The redshifts of the LBLs  cover the range $0.024-4.71$, whereas IBLs and HBLs are relatively nearby with $z<1$ and $z<1.2$, respectively. We note, however, that redshift estimations are not available for 8 sources in the sample (see Table \ref{listsources}). Fig.~\ref{isto}, shows the redshift distribution of all sources in the catalog.\\
\indent The spectral properties of each object have been derived using a power law and a logarithmic parabola to fit the spectra. In Fig.\ref{statistics} the distributions of Cstat/d.o.f for these models are shown for the different blazars sub-classes. These distributions overlap for the cases of LBLs and IBLs, while significant differences are present for HBL sources. This implies that for LBLs and IBLs, the power law is a good representation of the data and adding curvature does not improve the fit. Instead, for HBLs, the logarithmic parabola model provides a more appropriate description of the hard X-ray spectrum.\\
\indent Table~\ref{pofit} reports the best fit values of the parameters for the power-law while Table ~\ref{logparfit} contains the information derived using the logarithmic parabola model. In Fig.~\ref{gammaflux} the photon index from the power law model is plotted as a function of the observed flux in the 3-10 keV band for different blazar spectral classes. Such a phenomenological parameter clearly separates HBL blazars, which are characterised by steeper $\Gamma$s, than LBL sources. In fact, for HBLs and average photon index of $<\Gamma>$=2.56$\pm$0.30 is found, while for LBLs we computed a spectral slope of
$<\Gamma>$=1.58$\pm$0.22. Blazars with intermediate SEDs are mostly characterised by photon indices similar to those of LBLs, with a few exceptions.\\
\indent In Fig. ~\ref{beta} the curvature parameter ($\beta$) only for HBL sources is shown, since, as said, a power law model is a good representation for LBLs and IBLs sources (see Fig.\ref{statistics}). The spectral curvature derived for these sources can be either positive or negative suggesting strong variations in the source spectral shape. Interestingly, the curvature is not always correlated with the source flux and it seems to depend on each specific source, see bottom panel in Fig.~\ref{beta}. This behaviour is discussed in  Section \ref{soudetails}.\\
\indent The power law slope as a function of the broadband (3-30 keV) flux is shown in Fig. ~\ref{gamma1} for a subsample of multiply observed blazars. For some source (e.g. Mkn 501, 1ES 0229+200) these two quantities are anti-correlated, while no trends are observed for other objects (e.g. PKS 2155-304). Interestingly, this trend holds or not independently of the SED class of the source as for the case of the two HBLs Mkn 501 and PKS 2155-304. In Fig.~\ref{isto2}, we show the relation between the curvature parameter and the photon index for all the HBLs in the catalogue (shaded points). Moreover, we over plot the same parameters for the three blazars labelled, namely Mkn 421, Mkn 501 and PKS 2155-304. A weak and marginally significant anti-correlation is found between the spectral slope and the log parabola curvature (P$_{\rm cc}$=0.3 and P(<r)=0.04) for the whole sample of HBLs. However, we notice that that trend is driven by Mkn 421 data that are actually strongly anti-correlated. Once we do not consider data from this source, the the trend is even less significant with an compatible correlation coefficient and a null probability of P(<r)=8\%. We will further discuss these correlations for a selected sample of objects in the corresponding sub sections of Sect. 5.


\subsection{Sources observed multiple times}
\begin{figure}
	\centering
	\includegraphics[width=0.49\textwidth]{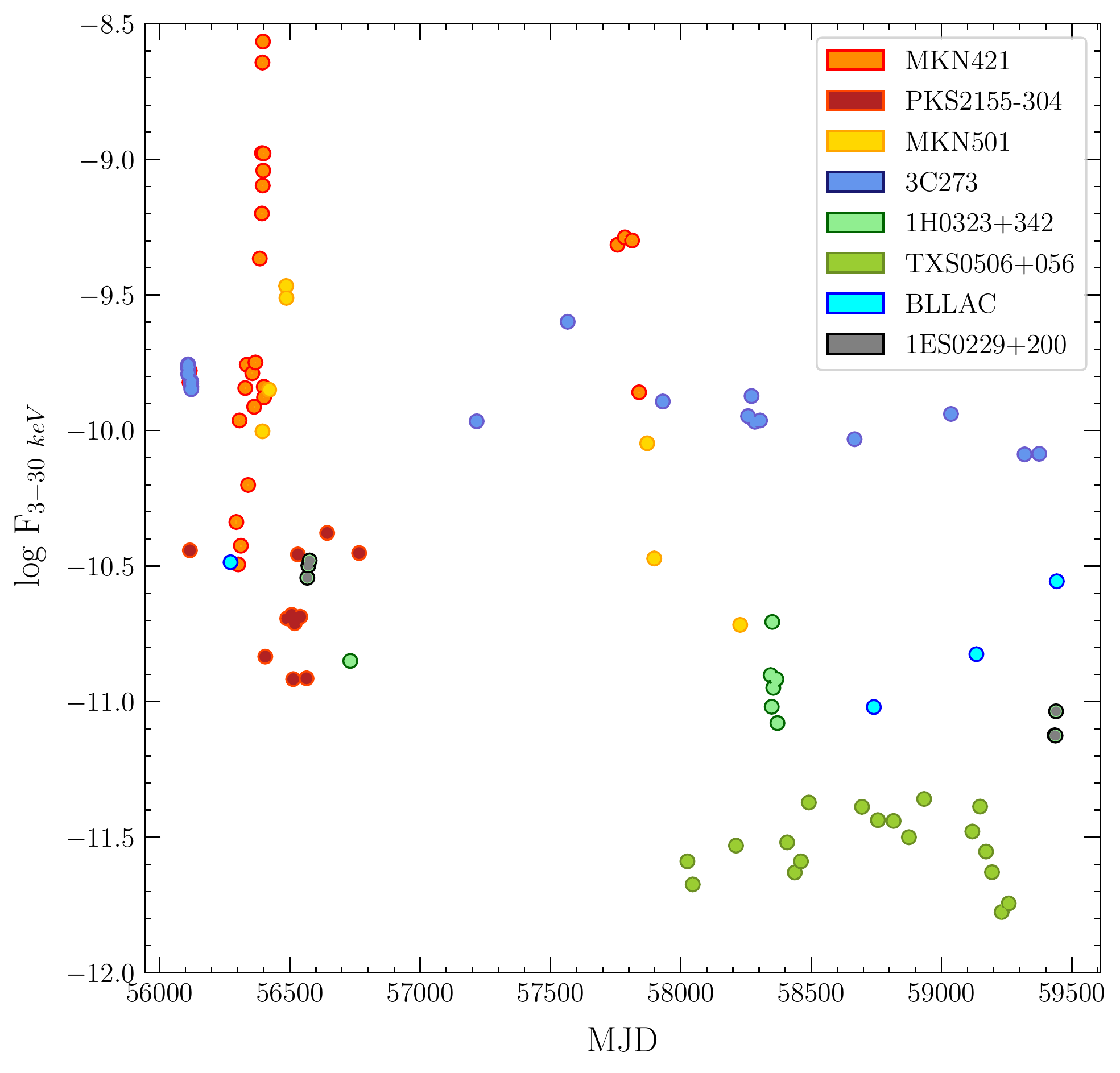}
	\caption{\small{Light curves of a selection of blazars observed several times. Errors at 68\% confidence level lie within the data points.}}\label{lcurves}
\end{figure}

 Thirty objects in our list were  observed by \N\, more than once (up to 26 times). Blazars are well known to show large amplitude flux changes at many frequencies and over a wide range of timescales. Our analysis confirms that the hard-X-ray energy band covered by \N\, is not an exception, as shown by Fig.~\ref{lcurves} where the flux in the 3-30 keV band is plotted against time. A variety of estimators have been used to perform variability characterisations such as the power-spectral density \citep[PSD.e.g.][]{Lawrence1993,Uttley2002,Uttley2005}, the fractional variability \citep[e.g.][]{Vaughan2003} and the structure function (SF) that has been used for ensemble studies \citep[e.g.][]{Vagn11,Vagn16,Midd17,Paolillo2017} or single source analyses \citep[e.g.][]{Gallo18}. The so-called normalised excess variance \citep[$\sigma^2_{\rm nxs}$, see e.g.][]{Ponti2012} is another tool that allows for variability estimates even when light curves are sparsely sampled.

The $\sigma^2_{\rm nxs}$ accounts for the mean quadratic variation corrected for the error, hence it provides a simple characterisation of the source intrinsic variability even for sources with only two flux measures at different epochs. This estimator is defined as $\sigma^2_{\rm nxs}$=(S$^2$-$\sigma_{\rm err}^2$)/<f>$^2$, where <f>$^2$ is the mean source flux computed over the available observations, S$^2$ is the light curve variance and the error term is the mean square photometric error associated to each flux. However, some caveats must be considered when using such an estimator. \cite{Allevato2013} pointed out that dealing with sparsely sampled light curves may underestimate of the variability itself due to its red-noise nature. On the other hand, \cite{Vagn16} urged caution on the use of this tool noting that cosmological time-dilation has to be accounted for and proposed a correction based on the ensemble properties of a statistically significant sample of radio quiet quasars. In other words, variability comparison of different sources must be performed over the same timescale. However, type 1 AGNs have a different driving emission mechanism, hence the correction by \cite{Vagn16} may not be straightforwardly applied since we lack quantitative information on the ensemble variability properties of blazars. Bearing this in mind, we report in Table \ref{nxs} the $\sigma^2_{\rm nxs}$. We notice that positive values for the $\sigma^2_{\rm nxs}$ were derived only for 9 sources, meaning that flux changes for the not-listed sources are compatible with the noise. The normalised excess variance in the 3-10 and 10-30 keV energy bands weights the actual amount of variability for each source. Interestingly, the normalised excess variance derived for LBL objects is compatible between the two bands while HBL show larger variations in the 10-30 keV energy range. This suggestive trend can result off the presence of a single (for LBL) or two (for HBL) emission components acting in the full 3-30 keV energy range.

\begin{table}
\centering
	\setlength{\tabcolsep}{1.5pt}
	\caption{\small{Normalised excess variance in the 3-10 and 10-30 keV energy bands computed in the source reference frame. The length of the light curves are showed in years.}\label{nxs}}
	\begin{tabular}{c c c c c c c}
		\hline
\hline
Source Name & $\sigma^2_{\rm nxs}$ (3-10 keV) &$\sigma^2_{\rm nxs}$ (10-30 keV)&$z$&$\Delta t$&Class&Obs\\
\hline
5BZBJ1104+3812 &1.55$\pm$ 0.01&2.23$\pm$ 0.02 &0.029 &4.58 &HBL&26\\
5BZQJ1229+0203 &0.07$\pm$ 0.01&0.09$\pm$ 0.01 &0.158 & 7.72 &LBL&25\\
5BZBJ2158-3013 &0.17$\pm$ 0.06&0.21$\pm$ 0.16 &0.116 & 1.10&HBL&9 \\
5BZBJ1653+3945 &0.75$\pm$ 0.03&0.96$\pm$ 0.05 &0.033 &4.86 &HBL&7 \\
5BZUJ0324+3410 &0.09$\pm$ 0.06&0.07$\pm$ 0.06 &0.061 &4.23 &IBL&7 \\
5BZGJ0232+2017 &0.36$\pm$ 0.09&0.51$\pm$ 0.14 &0.139 &6.90 &HBL&6 \\
5BZBJ2202+4216 &0.26$\pm$ 0.07&0.22$\pm$ 0.07 &0.069 &8.12 &IBL&4 \\
5BZQJ1256-0547 &0.25$\pm$ 0.07&0.31$\pm$ 0.08 &1.309 &3.17 &LBL&3 \\
5BZBJ0507+6737 &0.36$\pm$ 0.18&0.28$\pm$ 0.35 &0.416&0.11  &LBL&3 \\
\hline
\end{tabular}
\end{table}

\section{Spectral energy distributions and comments on selected objects}\label{soudetails}
The broad-band (radio to $\gamma$-ray) SED of all the sources in our sample have been assembled by 
combining the best fit spectral data derived in the previous sections with archival multi-frequency data, retrieved with the  VOU-Blazars tool \citep{Chang2019} and the SSDC SED Builder tool\footnote{https://tools.ssdc.asi.it/SED/}, as well as with the results of the Swift-XRT analysis of \cite{GiommiXRTspectra}. 
The Swift-XRT SED data of the sources not included in \cite{GiommiXRTspectra} have been  generated by us for this paper following the same procedure described in \cite{GiommiXRTspectra}. 
In the following we plot and describe the SED of a small number of selected objects, where the archival data appears as orange points, the Swift-XRT data as light green points, and the \NS\, data as magenta circles. The SEDs of all the 126 objects in the sample are available on-line at\\ {\url{https://openuniverse.asi.it/blazars/swift}}.

\subsection{Mkn 421}

The HBL source Mkn~421 is one of the best studied BL Lacs shining in the local Universe \citep[$z$=0.03,][]{deVaucouleurs1991}. Due to its brightness and proximity this source has been the focus of a number of multiwavelength campaigns resulting in several papers dissecting its specific properties \citep[e.g.][]{Acciari2020,Paliya2015,Kapanadze2016,Balokovic2016}. Mkn 421 shows flux and spectral variations across the entire electromagnetic spectrum, the most prominent ones being in the X-ray and very high energy $\gamma$-ray bands. The \N\ observations demonstrate that Mkn 421 is highly variable also in the hard X-ray band, as it is clear from Fig. \ref{lcurves} and Fig. \ref{mrksed} and from the remarkable normalised excess variance listed in Table \ref{nxs}. The brightest X-ray flare was observed on MJD 56396.921 when the 3-10 keV flux was $(15.16\pm0.02)\times10^{-10}\:{\rm erg\:cm^{-2}\:s^{-1}}$ which exceeds the lowest flux measured by \N\, by $\sim65$ times . In this period the source was in an active flaring state \cite[][]{Acciari2021} also in the Swift-XRT band \citep[see Fig. 11 of][]{GiommiXRTspectra}. Together with the flux, the photon index also varies following a harder-when-brighter trend. The hardest index of $2.280\pm0.003$ was estimated during the brightest X-ray flare. The SED of Mkn 421 is shown in Fig. \ref{mrksed}. It shows a clear trend of increase of the spectral curvature with flux. The curvature term $\beta$ of \textit{logpar} is well constrained in all the observations and switches from negative values to positive one (Fig. \ref{beta} lower panel). This parameter is strongly correlated with the source hard flux. We found indeed a P$_{\rm cc}$=0.75(0.73) with P(<r)$\sim$2(4)$\times$10$^{-5}$. In principle, this curvature could be due to the cooling or escape of high-energy electrons, i.e., as the peak of the low energy component is at the X-ray band ($10^{16}$ Hz), the radiative cooling time of the emitting electrons is of the order of one hour. Alternatively, the particle spectrum could be intrinsically curved. For example, \citet{2004A&A...413..489M} showed that logarithmic parabola spectra are naturally formed for an energy-dependent statistical acceleration.

\subsection{3C 273}

3C 273 is a nearby (z=0.158) high-luminosity FSRQ \citep{Strauss1992}, with an extended radio jet showing superluminal motion \citep{Seielstad1979,Pearson1981,Unwin1985}. Being a highly variable bright source at almost all wavelengths 3C\,273 has been monitored for many years with different instruments. \N\ observed this source 25 times between 2012 and 2021. The 2012 pointings were part of cross-calibration campaigns carried by the International Astronomical Consortium for High Energy Calibration (IACHEC) with Chandra, \N, \textit{Swift}, Suzaku, and XMM-Newton observatories on 3C 273 and PKS 2155-304 \cite[see][]{Madsen2017}. Combining the 2012 observations with data at other frequencies, \cite{Madsen2015A} showed that the \N\ spectrum cannot be fit simply by an absorbed power law, but the fit has residuals above $\sim$20 keV, thus they investigated the coronal emission and put constraints on a weak Compton reflection component in the hard X-ray band.
Fig. \ref{lcurves} shows the X-ray flux of 3C 273 in different snapshots, while the multiwavelength SED is shown in Fig. ~\ref{3c273sed}. The 3-10 keV flux does not show large variations with an average value of $\simeq6.51\times10^{-11}\:{\rm erg\:cm^{-2}\:s^{-1}}$; in the considered period, the highest and lowest flux differs only $3.1$ times. In our analysis we did not find any correlation between the photon index ($\Gamma$) and the flux, although $\Gamma$ changed significantly between different observations. During 2012, the photon index was mostly in the range $1.50-1.60$ while in later observations, it softened to values $>1.60$. The X-ray flux of 3C 273 versus the photon index is shown in Fig. \ref{gamma1}. In this X-ray range, the emission is due to the rising part of the inverse Compton component, which is defined by the power-law index of the emitting electrons. Such a change can be easily explained by related changes in the emitting electron spectrum.

\subsection{3C 279}

\begin{figure}
	\centering
	\includegraphics[width=0.49\textwidth]{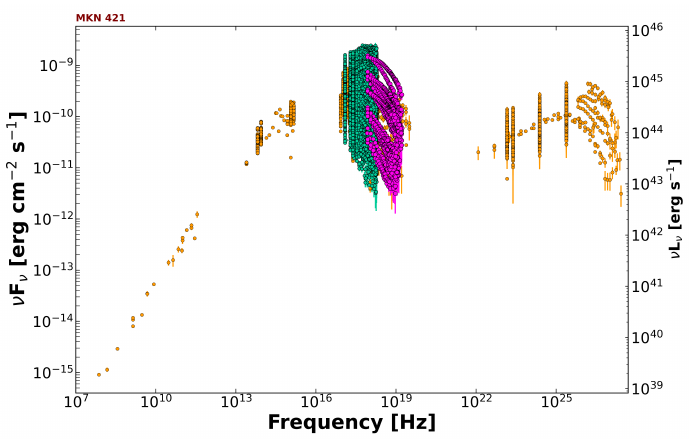}
	\caption{\small{The SED of Mkn 421 build with \N\,(magenta) Swift-XRT (green) and multi-frequency archival data (orange). The \N\, data are from our analysis, the Swift-XRT points are from \protect \cite{GiommiXRTspectra}, and the archival data have been retrieved using the VOU\_Blazars and the SSDC  SED Builder tools.}
	}\label{mrksed}
\end{figure}

 \begin{figure}
	\centering
	\includegraphics[width=0.49\textwidth]{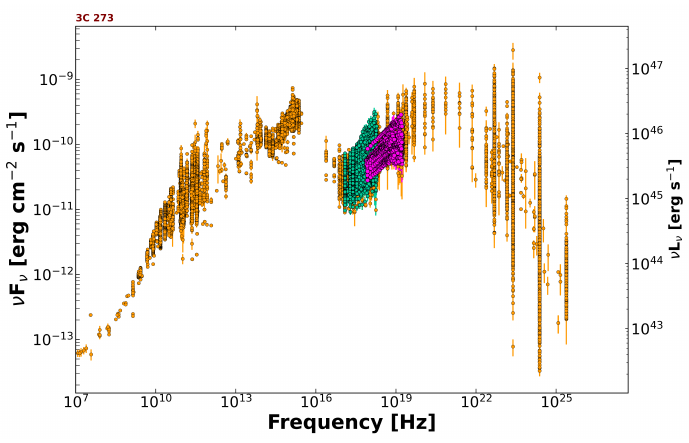}
	\caption{The SED of 3C 273. Data description and colour coding are the same as in Fig.\ref{mrksed}.
	}\label{3c273sed}
\end{figure}
\begin{figure}
	\centering
	\includegraphics[width=0.49\textwidth]{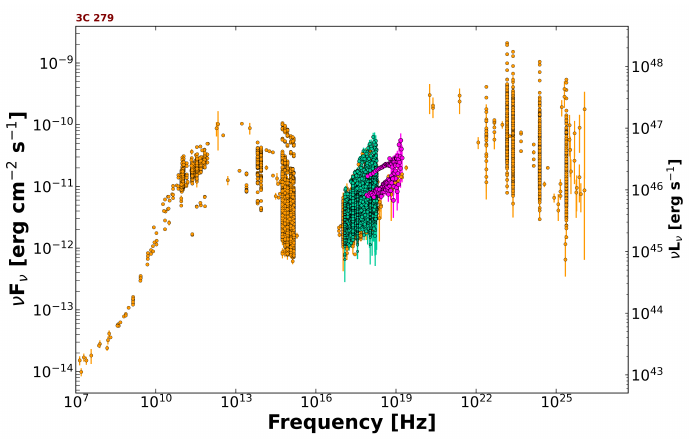}
	\caption{\small{The SED of 3C 279. Data description and colour coding are the same as in Fig.\ref{mrksed}.
	}}\label{3c279}
\end{figure}
\begin{figure}
	\centering
	\includegraphics[width=0.49\textwidth]{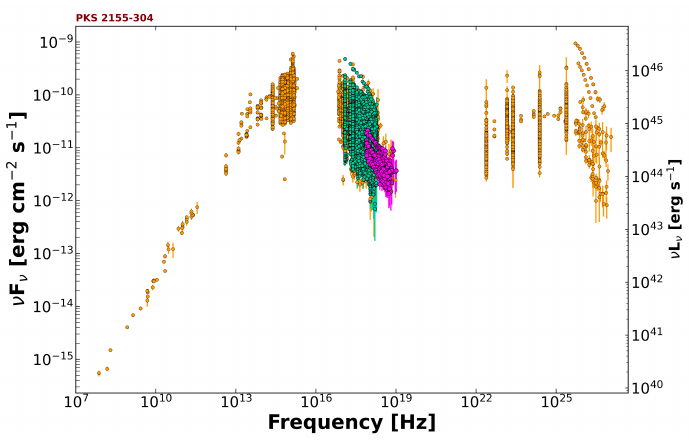}
	\caption{\small{The SED PKS 2155-304. Data description and colour coding are the same as in Fig.\ref{mrksed}.
	}}\label{pks2155sed}
\end{figure}

\begin{figure}
	\centering
	\includegraphics[width=0.49\textwidth]{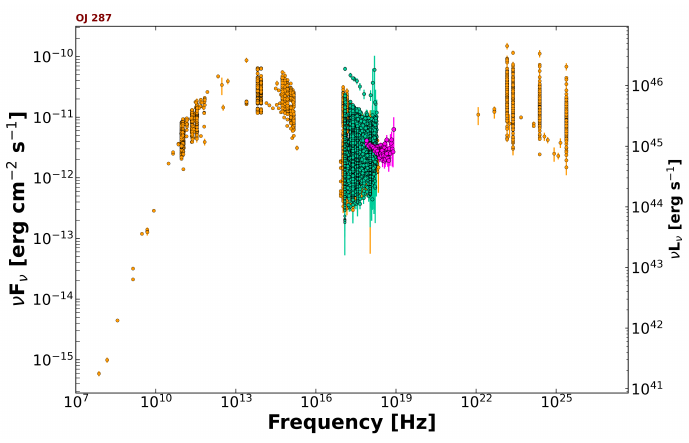}
	\caption{\small{The SED of 5BZBJ~0854+2006 also known as OJ 287.
	Data description and colour coding are the same as in Fig.\ref{mrksed}.
	}}\label{ojsed}
\end{figure}

\begin{figure}
	\centering
	\includegraphics[width=0.49\textwidth]{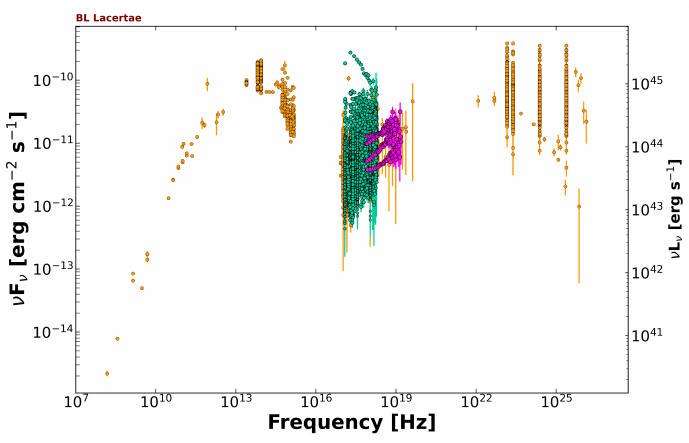}
	\caption{\small{The SED of BL Lacertae. 
	Data description and colour coding are the same as in Fig.\ref{mrksed}.
	}}\label{bllac}
\end{figure}

\begin{figure}
	\centering
	\includegraphics[width=0.49\textwidth]{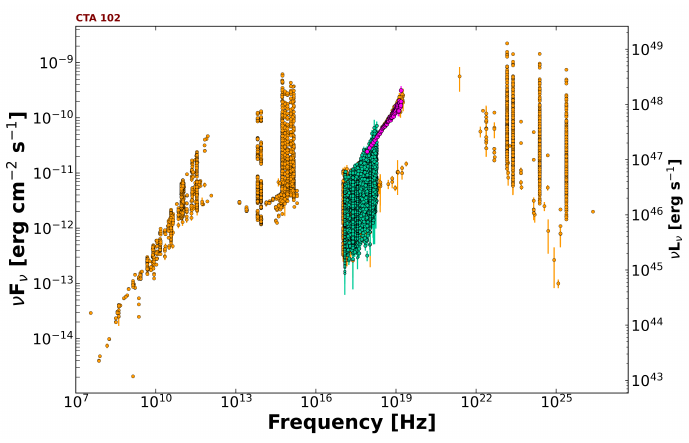}
	\caption{\small{The SED of CTA 102.
	Data description and colour coding are the same as in Fig.\ref{mrksed}.
	}}\label{ctased}
\end{figure}

3C 279 is a well-known FSRQ at $z=0.536$ \citep{Marziani1996}. \N\ observed three times this source, twice in 2014 and one time in 2021. 
The corresponding SED is shown in Fig~\ref{3c279}.
The best fit value of $\Gamma$ from our fits is $\leq1.90$, consistent with an LBL-like object where the hard X-rays are from SSC emission of synchrotron emitting electrons. Although three observations are not sufficient to attempt to find statistically meaningful correlations between different parameters, we can report that the 3.0-10 keV flux significantly varied from  $\sim8\times10^{-12}\:{\rm erg\:cm^{-2}\:s^{-1}}$ on MJD 56642.2 and 59317.1 to $\sim 1.85\times10^{-11}\:{\rm erg\:cm^{-2}\:s^{-1}}$ on MJD 566578. Based on Fermi-LAT data, \cite{Hayashida2015} reported on intra-day variability and a rapid $\gamma$-ray flare characterising the very high energy spectrum of this source. \cite{Da2015}, via SED modeling, attributed the X-ray emission to SSC, and the $\gamma$-ray emission to EC. This is particularly interesting as the origin of the hard X-ray emission in FSRQ-type blazars is still poorly understood \citep{Sikora2013}. In fact, FSRQs were not expected be bright TeV sources, until TeV observatories reported the discovery of strong and highly variable emission in a number of them, including 3C 279 \citep{MAGIC2008}. It has been then suggested that the TeV emission originates from a tail of high-energy electrons, whose synchrotron emission would appear in the hard X-ray band changing the spectral shape to softer slope, i.e. from $\Gamma \sim 1.5$ to $\Gamma \sim 2$ \citep[see e.g.][]{Donnarumma2013}. Therefore, it is particularly important to study in detail the hard X-ray emission from such sources; on the other hand, it is clear that \N\ data alone cannot totally unveil the complexity of 3C 279. Future observations, especially obtaining X-ray polarimetry data \citep{eXTP2019}, will help to better constrain the emission mechanism(s).

\subsection{PKS 2155-304}

PKS 2155-304 is a BL Lac at $z = 0.117$ \citep{Falomo1993}, detected well above the sensitivity limit at TeV energies \citep{Chad1999}. \citet[][]{Giommi2002babs} reported results from three BeppoSAX observations, during which the spectral curvature remained in a narrow range (0.27-0.3) while the flux changed from $2.5$ to $8.3 \times 10^{-11} {\rm erg~cm^{-2} s^{-1}}$. \cite{massarino2008} showed that in this source the X-ray emission is the high-energy end of the synchrotron emission, and that the SED peak location was falling outside the observational X-ray range. 
\N\, observed PKS 2155-304 nine times between 2012 and 2013. First results were reported by \cite{Madejski2016}. The source flux was varying between $(5.57\pm0.16)\times10^{-12}\:{\rm erg\:cm^{-2}\:s^{-1}}$ and $(1.67\pm0.15)\times10^{-11} {\rm erg\:cm^{-2}\:s^{-1}}$ in the 3-10 keV band.
As shown in Fig.\ref{gamma1}, this HBL object has a $\Gamma$ in the range 2.58-3.03 and does not show any evidence for a correlation between the photon index and the flux. On the other hand, the curvature parameter $\beta$ switched between negative and positive values and is correlated with the source flux (P$_{\rm cc}$=0.92 and P(<r)=0.0004), as in the case for Mkn 421, see Fig.~\ref{beta}. This may be due to a varying mix between the end of the synchrotron emission and the rise of the IC component. When the synchrotron component is bright and dominates the overall flux the spectrum is convex, whereas when the synchrotron emission is low the flux from the hard IC component is not negligible and the overall spectral curvature changes to a slightly concave shape.

\subsection{OJ 287}

OJ 287,  also known as 5BZB~J0854+2006,
is a well-studied blazar at $z = 0.306$. It is considered the best candidate AGN for
hosting a supermassive binary black hole \citep[see e.g.][]{Britzen2018}. Recently, \cite{Meyer2019} described a
low-state Fermi-LAT $\gamma$-ray spectrum associated with the kpc scale jet of OJ 287 that is consistent with the
predictions of the IC/CMB model for their X-ray emission. However, \cite{Pal2020} reported a strong soft X-ray excess in
the X-ray spectrum of OJ287 using XMM-Newton observations. Furthermore, \cite{Lange2020} showed that the
\textit{Swift}-BAT spectrum (20-100 keV) is well-described by a power-law with $\Gamma = 1.31$, using the 105-months
\textit{Swift}-BAT survey. \N\ observed OJ287 twice during two major flare events detected by Swift (in 2017 and 2020): although the 3.0-10.0 keV X-ray flux was similar 
($\simeq3.5\times10^{-12}\:{\rm erg\:cm^{-2}\:s^{-1}}$) during these observations, the photon index varied. It was $2.03\pm0.04$ in 2017 and $2.42\pm0.06$ in 2020. These photon indices are significantly different than those estimated in
the 0.1–50 keV band by BeppoSAX: in fact, these observations can be fitted to a power-law with $\Gamma = 1.95^{+0.30}_{-0.29}$ (24-Nov-1997) and $\Gamma =1.64^{+0.22}_{-0.20}$ \cite[20-Nov-2001, see Tab. 3 in][]{Don2005}. 
Therefore the observations by BeppoSAX and \N\ were probing probably two different source states. The observed multiwavelength SED during the flare activities of OJ 287 have been attributed to the emerging of a HBL-like broadband
non-thermal emission component  due to the tail of high-energy electrons generating synchrotron and dominating
over IC in the hard X-ray band \cite[see e.g.][]{Komossa2020,Kushwaha2020}. The combination of Swift-XRT and \N\, observations shown in Fig. \ref{ojsed} (green and magenta points respectively) show an X-ray spectrum that is dominated by a flat IC component during faint states and by the steep tail of the synchrotron component during high states. The two \N\, observations, which were performed when the source was in intermediate intensity state, clearly caught both components. 

\subsection{BL Lacertae}

BL Lacertae, the prototype of the class of BL Lac objects, is a remarkable blazar not only for historical reasons and for its brightness, but also because it occasionally shows the presence of a broad H-alpha line \citep{1996MNRAS.281..737C, 2010A&A...516A..59C}, temporarily 
contradicting the definition of BL Lac objects as sources with featureless optical spectrum. The Swift-XRT data of this object also shows a widely changing behaviour (green points in Fig.\ref{bllac}), with a typically hard and variable X-ray spectrum, but also large soft X-ray flares (the largest one recorded in late 2020) during which the X-ray slope becomes much steeper, likely as the result of the synchrotron peak shifting to higher energies, turning this source from an IBL into a HBL blazar. 
The HBL component was interpreted as due to the emergence of synchrotron emission from freshly accelerated particles in a second emission zone \citep{BLLAC2021}. {\it NuSTAR} observed BL Lac three times \citep{Rani2017}, in all cases detecting a hard X-ray spectrum, with a further flattening above 10 keV in at least two occasions (magenta points in Fig. \ref{bllac}). One of the \textit{NuStar} observations was carried out five days after the large flare of 2020. The corresponding spectrum is the one with intermediate intensity and sharply hardening slope at high energies (see Fig.\ref{bllac}), probably indicating the on-set of a second, much harder component, detected also during the faintest state, independent of the softer ($\sim$ 1-10 keV) X-ray emission and possibly related to the $\gamma$-ray emission. Detailed modeling of BL Lacertae SEDs between 2008 and 2021, including the periods of \N\ observations, is presented in \citet{BLLAC2021}.

\subsection{CTA 102}

CTA 102 is among the brightest FSRQs in the high energy $\gamma$-ray band despite its large distance, at z = 1.037. Fermi-LAT observations revealed that the flux of this blazar sometimes exceeds $10^{-5}\: {\rm photon\:cm^{-2}\: s^{-1}}$ \cite[][]{Gasp2018}. CTA~102 was observed by \N\ only once in 2016, during a major flare when it was also bright in $\gamma$-rays. As shown by the SED of Fig. \ref{ctased}, \N\ data probe the rise of the IC component at its maximum  state ever observed in the hard X-rays. Our result ($\Gamma \sim 1.3$) is fully consistent with an earlier work \cite[][]{Gasp2018}, which also showed that \textit{Swift} monitoring detected a harder when brighter behaviour. The broadband SED of CTA 102, including the \N\ data (see Fig. \ref{ctased}) was modeled assuming that the jet dissipation occurs close (within BLR) or far from (outside BLR) the central source. The SED of CTA 102 from X-ray to $\gamma$-ray bands was modeled considering the IC scattering of synchrotron, or BLR reflected, or torus photons \citep{2020A&A...635A..25S}.

\begin{figure}
	\centering
	\includegraphics[width=0.49\textwidth]{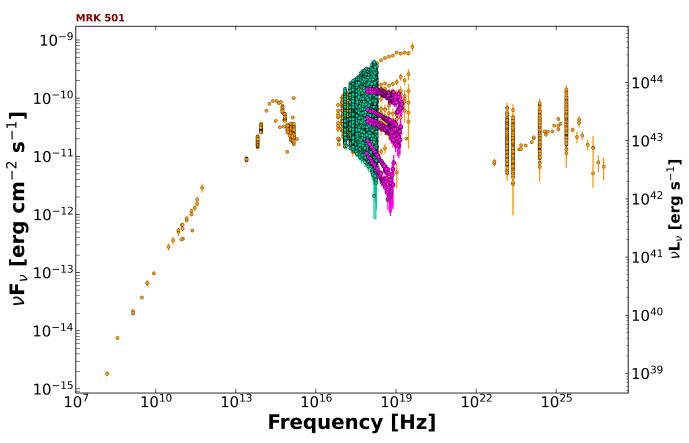}
	\caption{\small{The SED of Mkn 501.
	Data description and colour coding are the same as in Fig.\ref{mrksed}.
	}}\label{mkn501sed}
\end{figure}

\begin{figure}
	\centering
	\includegraphics[width=0.49\textwidth]{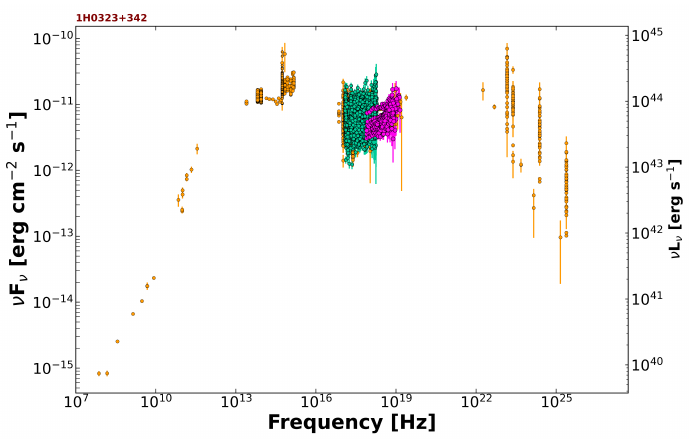}
	\caption{\small{The SED of 1H 0323+342.
	Data description and colour coding are the same as in Fig.\ref{mrksed}.
	}}\label{1h0323_sed}
\end{figure}

\begin{figure}
	\centering
	\includegraphics[width=0.49\textwidth]{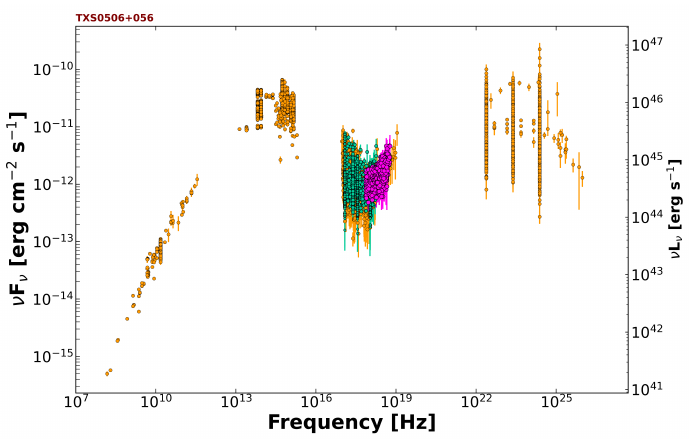}
	\caption{\small{The SED of TXS0506+056. 
	Data description and colour coding are the same as in Fig.\ref{mrksed}.
	}}\label{5BZBJ0509_sed}
\end{figure}

\subsection{Mkn 501}

Mkn 501, a nearby ($z=0.033$) HBL blazar \citep{deVaucouleurs1991} characterised by one the highest observed \nup\, values \citep{pian1998} and among the first BL Lac objects detected at TeV energies \citep{Weekes1996,whipple1999,Quinn1999}. 
Its SED and variability properties have been widely studied based on many observations, including \N\, data and on  multi-wavelength campaigns \cite[e.g.][]{Furniss2015,Swift2020}. \N\ observed this source 7 times between 2013 and 2018. Our analysis shows that as in the case of Mkn 421, Mkn 501 exhibits a typical harder when brighter behaviour. This behaviour is also evident in Fig. ~\ref{gamma1}, which shows that the photon index is strongly anti-correlated with the 3-30 keV band flux (P$_{\rm cc}$=-0.95,P(<r)=0.001). The hardest index of $2.08\pm0.01$ was observed on MJD 56485.91 when the 3.0-10 keV flux was $(1.56\pm0.07)\times10^{-10}\:{\rm erg\:cm^{-2}\:s^{-1}}$. Interestingly, the curvature term $\beta$ of \textit{logpar} is well constrained in all the observations, and consistent with 0 in May 2017. However, $\beta$ does not seem to be correlated with the flux, at variance with Mkn 421. From the SED of Fig.~\ref{mkn501sed} we can infer that \N\ observations trace the high energy tail of the synchrotron emission.\\

\subsection{1H 0323+342}

1H 0323+342 is a nearby source \citep[$z=0.061$,][]{Marcha1996}, characterized by a flat radio spectrum and an IBL-like SED. It is listed among the blazars of uncertain type in the BZCAT catalog, however it has been classified by other authors as a $\gamma$-ray emitting narrow-line Seyfert 1 ($\gamma$NLS1), a rare class of AGNs with low black hole masses \cite[see e.g.][]{Paliya2014}. In these sources, the jet emission contributes to the hard X-ray radiation, resulting in power-law dominated spectra. \cite{Kynoch2018} found that the broadband SED of 1H 0323+342 shows a high Compton dominance, as for FSRQs, but the luminosity is similar to a BL Lac. Furthermore, they reported the detection of a weak iron line feature with XMM-Newton. The multiwavelength SED of 1H 0323+342 and the X-ray (Swift-XRT) and $\gamma$-ray (Fermi-LAT) light curves are presented in \citet{2018IJMPD..2744001B}. \N\ observed this source 7 times (once in 2014 and the remaining 6 times in 2018). In these observations, the flux was found to be mostly at the level of $(4-6)\times10^{-12}\:{\rm erg\:cm^{-2}\:s^{-1}}$ and only on MJD 58350.62 it increased up to $(9.13\pm0.03)\times10^{-12}\:{\rm erg\:cm^{-2}\:s^{-1}}$. The photon index remained approximately constant ($\Gamma \sim 1.8$) in all observations. Recently, \cite{Mundo2020} presented X-ray spectral and timing analyses using all \N\ data together with simultaneous \textit{XMM-Newton} coverage, showing that the X-ray emission in the 0.5-79 keV range (soft excess and iron line) can be described with a combination of a black body component and relativistic reflection model. They also found a hard excess at high energies, which can be explained as the contribution of non-thermal jet emission to the hard X-ray spectrum. Fig.~\ref{1h0323_sed} shows the SED of 1H 0323+342.

\subsection{TXS 0506+056}

TXS 0506+056 is a relatively bright BL Lac object with an IBL-like SED. This blazar became famous and widely studied after its association in September 2017 with the high-energy neutrino IceCube-170922A \citep[][]{IceCube,neutrino,Dissecting,Ansoldi2018}. Before the occurrence of this event no redshift estimation was available for this object.
Shortly afterward \cite{Paiano2018} obtained a high quality optical spectrum at the 10.4 m Gran Telescopio Canarias, which revealed very weak emission lines at the redshift of $z=0.3365$. \cite{Padovani2019} showed that this source is in fact a masquerading BL Lac, that is a blazar that is intrinsically a FSRQ where the disk and broad-line emission is present but completely diluted by the strong non-thermal radiation from the jet. As part of the intense follow-up campaign aimed at studying in detail the likely electromagnetic counterpart to IceCube-170922A \N\ observed this blazar 18 times between 2017 and 2021. Our analysis does not show any correlation between different parameters, e.g. $\Gamma$ versus flux. The photon index is hard with a mean of $1.67$ and does not changes substantially in different observations. The 3.0-10 keV X-ray flux is also relatively constant with only about a factor 2 difference between the highest and lowest states. The SED of this source is plotted in Fig.~\ref{5BZBJ0509_sed}. The origin of multiwavelength emission from this sources has been widely discussed in the literature where leptonic, hadronic and lepto-hadronic models have been considered \citep[e.g., see ][]{Ansoldi2018,2018ApJ...864...84K, 2018ApJ...865..124M, 2018ApJ...866..109S, 2019MNRAS.484.2067R,2019MNRAS.483L..12C, 2019NatAs...3...88G, 2022MNRAS.509.2102G}.

\section{Final comments}

\indent We have compiled \textit{NuBlazar}, the first catalogue of hard X-ray spectra and spectral fit results through a standard processing of all the observations of the blazars included in the \N\, public archive as of September 2021. This database, which is analogous to the one presented by \cite{GiommiXRTspectra} for the sample of blazars frequently observed by Swift-XRT, includes 253 observations of 126 individual sources. The spectral data, along with other multi-frequency archival data sets, such as those that can be retrieved via e.g. the VOU\_Blazar tool and the SSDC SED builder, can be combined to build time-dependent SEDs and can be used for various purposes, including detailed physical model fitting. One such example is the recent work of \cite{BLLAC2021}, which fitted blazar models to 13 years of broad-band multi-frequency observations of BL Lacertae, the prototype of the class of blazars.\\
\indent The sample of blazars observed by \N\, is heterogeneous as it is the result of a wide range of uncorrelated scientific projects carried out over the years as part of the guest observers and other \N\, observation programs. As such it does not accurately represent the underlying population of blazars; however it includes most of the well known objects, and it is sufficiently large to encompass blazars of all types.\\
\indent The SEDs of the sources in the sample, some of which are shown from Fig. \ref{mrksed} to Fig. \ref{5BZBJ0509_sed} confirm that \N\, probes the steep end of the very variable synchrotron emission in most HBL sources and of the hard Inverse Compton radiation in IBL and LBL blazars. In some cases, like OJ~287, both components have been detected (see Fig. \ref{ojsed}). Within the statistical limits of the sample we note that the X-ray spectral slope in the \N\ energy range is a good predictor of the SED type of blazars independently of their flux. This is clear from Fig. \ref{gammaflux} where HBLs  and LBLs, are well separated, while  IBLs are mostly located in the LBL area with only a few  exceptions that occur when \nup\, moves to higher values.\\
\indent Another well known characteristics of blazars' X-ray spectra is the presence of curvature, usually represented by a log parabola model \citep{Massaro2004}. In the \N\, energy band our results show that this feature is significantly detected (see Figs. \ref{statistics}) only in bright HBL sources where its amount depends on flux level (Fig. \ref{beta}), likely reflecting the known correlation between \nup\, and source intensity \citep{GiommiXRTspectra}.\\
\indent In closing, \N~data had proven to be extremely suitable to investigate blazars properties (e.g. $\Gamma$, $\alpha$ and $\beta$), thus they are fundamental complement for currently operating X-ray mission such as IXPE \cite[][]{Weisskopf2016} that will provide the first spectro-polarimetric data.



\section{ACKNOWLEDGEMENTS}

The authors thank the referee for his/her useful comments to the paper. \textbf{RM} thanks Fausto Vagnetti for discussions and acknowledges financial support of INAF (Istituto Nazionale di Astrofisica), Osservatorio Astronomico di Roma, ASI (Agenzia Spaziale Italiana) under contract to INAF: ASI 2014-049-R.0 dedicated to SSDC. We acknowledge the use of data, analysis tools and services from the Open Universe platform, the ASI Space Science Data Center (SSDC), the Astrophysics Science Archive Research Center (HEASARC), the Astrophysics Data System (ADS), and the National Extra-galactic Database (NED). This  work  is  based  on  observations obtained with: the NuSTAR mission,  a  project  led  by  the  California  Institute  of  Technology,  managed  by  the  Jet  Propulsion  Laboratory  and  funded  by  NASA. \textbf{NS} acknowledges the support by the Science Committee of RA, in the frames of the research projects No 20TTCG-1C015 and 21T-1C260.

\section{Data Availability}
In agreement with our previous works based on the \textit{Swift\_deepsky}  \citep[][]{Giommi2019} and the \textit{Swift\_xrtproc} software tools \citep{GiommiXRTspectra}, the \NS\, pipeline used in this work are made available as part of the OU and ASI Space Science Data Center's activities. All spectral fits products and SEDs of the sources observed by \N, generated as part of this work, are available either through the the SSDC web tools \footnote{\url{https://www.ssdc.asi.it/}} or within the OU platform in different forms, such as: 
\begin{itemize}
\item a dedicated SSDC interactive web table at  \url{https://www.ssdc.asi.it/nustarblaz/}
\item through the VOU-Blazars and other tools within the OU portal \footnote{\url{https://openuniverse.asi.it}} 
\item through the BSDC VO query interface \url{http://vo.bsdc.icranet.org/}
\item through the SSDC SED builder tool \footnote{\url{https://tools.ssdc.asi.it/SED/}}.
\end{itemize}

\bibliographystyle{mnras}

\begin{thebibliography}{}
\makeatletter
\relax
\def\mn@urlcharsother{\let\do\@makeother \do\$\do\&\do\#\do\^\do\_\do\%\do\~}
\def\mn@doi{\begingroup\mn@urlcharsother \@ifnextchar [ {\mn@doi@}
  {\mn@doi@[]}}
\def\mn@doi@[#1]#2{\def\@tempa{#1}\ifx\@tempa\@empty \href
  {http://dx.doi.org/#2} {doi:#2}\else \href {http://dx.doi.org/#2} {#1}\fi
  \endgroup}
\def\mn@eprint#1#2{\mn@eprint@#1:#2::\@nil}
\def\mn@eprint@arXiv#1{\href {http://arxiv.org/abs/#1} {{\tt arXiv:#1}}}
\def\mn@eprint@dblp#1{\href {http://dblp.uni-trier.de/rec/bibtex/#1.xml}
  {dblp:#1}}
\def\mn@eprint@#1:#2:#3:#4\@nil{\def\@tempa {#1}\def\@tempb {#2}\def\@tempc
  {#3}\ifx \@tempc \@empty \let \@tempc \@tempb \let \@tempb \@tempa \fi \ifx
  \@tempb \@empty \def\@tempb {arXiv}\fi \@ifundefined
  {mn@eprint@\@tempb}{\@tempb:\@tempc}{\expandafter \expandafter \csname
  mn@eprint@\@tempb\endcsname \expandafter{\@tempc}}}

\bibitem[\protect\citeauthoryear{Abdo et~al.,}{Abdo et~al.}{2010}]{Abdo2010}
Abdo A.,  et~al., 2010, \mn@doi [ApJ] {10.1088/0004-637X/716/1/30}, 716, 30

\bibitem[\protect\citeauthoryear{{Acciari} et~al.,}{{Acciari}
  et~al.}{2020}]{Acciari2020}
{Acciari} V.~A.,  et~al., 2020, \mn@doi [\apjs] {10.3847/1538-4365/ab89b5},
  \href {https://ui.adsabs.harvard.edu/abs/2020ApJS..248...29A} {248, 29}

\bibitem[\protect\citeauthoryear{{Acciari} et~al.,}{{Acciari}
  et~al.}{2021}]{Acciari2021}
{Acciari} V.~A.,  et~al., 2021, \mn@doi [\mnras] {10.1093/mnras/staa3727},
  \href {https://ui.adsabs.harvard.edu/abs/2021MNRAS.504.1427A} {504, 1427}

\bibitem[\protect\citeauthoryear{{Ackermann} et~al.,}{{Ackermann}
  et~al.}{2015a}]{Ackermann2015Cat}
{Ackermann} M.,  et~al., 2015a, VizieR Online Data Catalog, \href
  {https://ui.adsabs.harvard.edu/abs/2015yCat..18100014A} {p. J/ApJ/810/14}

\bibitem[\protect\citeauthoryear{{Ackermann} et~al.,}{{Ackermann}
  et~al.}{2015b}]{Ackermann2015}
{Ackermann} M.,  et~al., 2015b, \mn@doi [\apj] {10.1088/0004-637X/810/1/14},
  \href {https://ui.adsabs.harvard.edu/abs/2015ApJ...810...14A} {810, 14}

\bibitem[\protect\citeauthoryear{{Aharonian}}{{Aharonian}}{2000}]{Aharonian2000}
{Aharonian} F.~A.,  2000, \mn@doi [\na] {10.1016/S1384-1076(00)00039-7}, \href
  {https://ui.adsabs.harvard.edu/abs/2000NewA....5..377A} {5, 377}

\bibitem[\protect\citeauthoryear{{Ajello} et~al.,}{{Ajello}
  et~al.}{2014}]{ajello2014}
{Ajello} M.,  et~al., 2014, \mn@doi [\apj] {10.1088/0004-637X/780/1/73}, \href
  {http://adsabs.harvard.edu/abs/2014ApJ...780...73A} {780, 73}

\bibitem[\protect\citeauthoryear{{Allevato}, {Paolillo}, {Papadakis}  \&
  {Pinto}}{{Allevato} et~al.}{2013}]{Allevato2013}
{Allevato} V.,  {Paolillo} M.,  {Papadakis} I.,   {Pinto} C.,  2013, \mn@doi
  [\apj] {10.1088/0004-637X/771/1/9}, \href
  {https://ui.adsabs.harvard.edu/abs/2013ApJ...771....9A} {771, 9}

\bibitem[\protect\citeauthoryear{{Ansoldi} et~al.,}{{Ansoldi}
  et~al.}{2018}]{Ansoldi2018}
{Ansoldi} S.,  et~al., 2018, \mn@doi [\apjl] {10.3847/2041-8213/aad083}, \href
  {https://ui.adsabs.harvard.edu/abs/2018ApJ...863L..10A} {863, L10}

\bibitem[\protect\citeauthoryear{{Arnaud}}{{Arnaud}}{1996}]{Arna96}
{Arnaud} K.~A.,  1996, in {Jacoby} G.~H.,  {Barnes} J.,  eds,  Astronomical
  Society of the Pacific Conference Series Vol. 101, Astronomical Data Analysis
  Software and Systems V. p.~17

\bibitem[\protect\citeauthoryear{{Baghmanyan} \& {Sahakyan}}{{Baghmanyan} \&
  {Sahakyan}}{2018}]{2018IJMPD..2744001B}
{Baghmanyan} V.,  {Sahakyan} N.,  2018, \mn@doi [International Journal of
  Modern Physics D] {10.1142/S0218271818440017}, \href
  {https://ui.adsabs.harvard.edu/abs/2018IJMPD..2744001B} {27, 1844001}

\bibitem[\protect\citeauthoryear{{Balokovi{\'c}} et~al.,}{{Balokovi{\'c}}
  et~al.}{2016}]{Balokovic2016}
{Balokovi{\'c}} M.,  et~al., 2016, \mn@doi [\apj]
  {10.3847/0004-637X/819/2/156}, \href
  {https://ui.adsabs.harvard.edu/abs/2016ApJ...819..156B} {819, 156}

\bibitem[\protect\citeauthoryear{{Bhatta}, {Mohorian}  \& {Bilinsky}}{{Bhatta}
  et~al.}{2018}]{Bhatta2018}
{Bhatta} G.,  {Mohorian} M.,   {Bilinsky} I.,  2018, \mn@doi [\aap]
  {10.1051/0004-6361/201833628}, \href
  {https://ui.adsabs.harvard.edu/abs/2018A&A...619A..93B} {619, A93}

\bibitem[\protect\citeauthoryear{{Bianchi}, {Maiolino}  \&
  {Risaliti}}{{Bianchi} et~al.}{2012}]{Bianchi2012}
{Bianchi} S.,  {Maiolino} R.,   {Risaliti} G.,  2012, \mn@doi [Advances in
  Astronomy] {10.1155/2012/782030}, \href
  {https://ui.adsabs.harvard.edu/abs/2012AdAst2012E..17B} {2012, 782030}

\bibitem[\protect\citeauthoryear{{Blackburn}}{{Blackburn}}{1995}]{Blackburn1995}
{Blackburn} J.~K.,  1995, in {Shaw} R.~A.,  {Payne} H.~E.,   {Hayes} J.~J.~E.,
  eds,  {Astronomical Society of the Pacific Conference Series} Vol. 77,
  {Astronomical Data Analysis Software and Systems IV}. p.~367

\bibitem[\protect\citeauthoryear{{B{\l}a{\.z}ejowski}, {Sikora}, {Moderski}  \&
  {Madejski}}{{B{\l}a{\.z}ejowski} et~al.}{2000}]{Blazejowski2000}
{B{\l}a{\.z}ejowski} M.,  {Sikora} M.,  {Moderski} R.,   {Madejski} G.~M.,
  2000, \mn@doi [\apj] {10.1086/317791}, \href
  {https://ui.adsabs.harvard.edu/abs/2000ApJ...545..107B} {545, 107}

\bibitem[\protect\citeauthoryear{{B{\"o}ttcher}, {Reimer}, {Sweeney}  \&
  {Prakash}}{{B{\"o}ttcher} et~al.}{2013}]{Bo2013}
{B{\"o}ttcher} M.,  {Reimer} A.,  {Sweeney} K.,   {Prakash} A.,  2013, \mn@doi
  [\apj] {10.1088/0004-637X/768/1/54}, \href
  {https://ui.adsabs.harvard.edu/abs/2013ApJ...768...54B} {768, 54}

\bibitem[\protect\citeauthoryear{{Britzen} et~al.,}{{Britzen}
  et~al.}{2018}]{Britzen2018}
{Britzen} S.,  et~al., 2018, \mn@doi [\mnras] {10.1093/mnras/sty1026}, \href
  {https://ui.adsabs.harvard.edu/abs/2018MNRAS.478.3199B} {478, 3199}

\bibitem[\protect\citeauthoryear{{Capetti}, {Raiteri}  \&
  {Buttiglione}}{{Capetti} et~al.}{2010}]{2010A&A...516A..59C}
{Capetti} A.,  {Raiteri} C.~M.,   {Buttiglione} S.,  2010, \mn@doi [\aap]
  {10.1051/0004-6361/201014232}, \href
  {https://ui.adsabs.harvard.edu/abs/2010A&A...516A..59C} {516, A59}

\bibitem[\protect\citeauthoryear{{Cash}}{{Cash}}{1979}]{Cash79}
{Cash} W.,  1979, \mn@doi [\apj] {10.1086/156922}, \href
  {http://adsabs.harvard.edu/abs/1979ApJ...228..939C} {228, 939}

\bibitem[\protect\citeauthoryear{{Cerruti}, {Zech}, {Boisson}, {Emery}, {Inoue}
   \& {Lenain}}{{Cerruti} et~al.}{2019}]{2019MNRAS.483L..12C}
{Cerruti} M.,  {Zech} A.,  {Boisson} C.,  {Emery} G.,  {Inoue} S.,   {Lenain}
  J.-P.,  2019, \mn@doi [\mnras] {10.1093/mnrasl/sly210}, \href
  {https://ui.adsabs.harvard.edu/abs/2019MNRAS.483L..12C} {483, L12}

\bibitem[\protect\citeauthoryear{{Chadwick} et~al.,}{{Chadwick}
  et~al.}{1999}]{Chad1999}
{Chadwick} P.~M.,  et~al., 1999, \mn@doi [\apj] {10.1086/306862}, \href
  {https://ui.adsabs.harvard.edu/abs/1999ApJ...513..161C} {513, 161}

\bibitem[\protect\citeauthoryear{{Chang}, {Arsioli}, {Giommi}, {Padovani}  \&
  {Brandt}}{{Chang} et~al.}{2019a}]{Chang2019Cat}
{Chang} Y.~L.,  {Arsioli} B.,  {Giommi} P.,  {Padovani} P.,   {Brandt} C.,
  2019a, VizieR Online Data Catalog, \href
  {https://ui.adsabs.harvard.edu/abs/2019yCat..36320077C} {pp J/A+A/632/A77}

\bibitem[\protect\citeauthoryear{{Chang}, {Arsioli}, {Giommi}, {Padovani}  \&
  {Brandt}}{{Chang} et~al.}{2019b}]{Chang2019}
{Chang} Y.~L.,  {Arsioli} B.,  {Giommi} P.,  {Padovani} P.,   {Brandt} C.~H.,
  2019b, \mn@doi [\aap] {10.1051/0004-6361/201834526}, \href
  {https://ui.adsabs.harvard.edu/abs/2019A&A...632A..77C} {632, A77}

\bibitem[\protect\citeauthoryear{{Corbett}, {Robinson}, {Axon}, {Hough},
  {Jeffries}, {Thurston}  \& {Young}}{{Corbett}
  et~al.}{1996}]{1996MNRAS.281..737C}
{Corbett} E.~A.,  {Robinson} A.,  {Axon} D.~J.,  {Hough} J.~H.,  {Jeffries}
  R.~D.,  {Thurston} M.~R.,   {Young} S.,  1996, \mn@doi [\mnras]
  {10.1093/mnras/281.3.737}, \href
  {https://ui.adsabs.harvard.edu/abs/1996MNRAS.281..737C} {281, 737}

\bibitem[\protect\citeauthoryear{{Dermer} \& {Schlickeiser}}{{Dermer} \&
  {Schlickeiser}}{1993}]{Dermer93}
{Dermer} C.~D.,  {Schlickeiser} R.,  1993, \mn@doi [\apj] {10.1086/173251},
  \href {https://ui.adsabs.harvard.edu/abs/1993ApJ...416..458D} {416, 458}

\bibitem[\protect\citeauthoryear{{Donato}, {Sambruna}  \& {Gliozzi}}{{Donato}
  et~al.}{2005a}]{Donato2005}
{Donato} D.,  {Sambruna} R.~M.,   {Gliozzi} M.,  2005a, \mn@doi [\aap]
  {10.1051/0004-6361:20034555}, \href
  {https://ui.adsabs.harvard.edu/abs/2005A&A...433.1163D} {433, 1163}

\bibitem[\protect\citeauthoryear{{Donato}, {Sambruna}  \& {Gliozzi}}{{Donato}
  et~al.}{2005b}]{Don2005}
{Donato} D.,  {Sambruna} R.~M.,   {Gliozzi} M.,  2005b, \mn@doi [\aap]
  {10.1051/0004-6361:20034555}, \href
  {https://ui.adsabs.harvard.edu/abs/2005A&A...433.1163D} {433, 1163}

\bibitem[\protect\citeauthoryear{{Donnarumma}, {Tramacere}, {Turriziani},
  {Costamante}, {Campana}, {De Rosa}  \& {Bozzo}}{{Donnarumma}
  et~al.}{2013}]{Donnarumma2013}
{Donnarumma} I.,  {Tramacere} A.,  {Turriziani} S.,  {Costamante} L.,
  {Campana} R.,  {De Rosa} A.,   {Bozzo} E.,  2013, in European Physical
  Journal Web of Conferences. p. 04015 (\mn@eprint {arXiv} {1310.6965}),
  \mn@doi{10.1051/epjconf/20136104015}

\bibitem[\protect\citeauthoryear{{Falomo}, {Pesce}  \& {Treves}}{{Falomo}
  et~al.}{1993}]{Falomo1993}
{Falomo} R.,  {Pesce} J.~E.,   {Treves} A.,  1993, \mn@doi [\apjl]
  {10.1086/186913}, \href
  {https://ui.adsabs.harvard.edu/abs/1993ApJ...411L..63F} {411, L63}

\bibitem[\protect\citeauthoryear{{Furniss}, {Noda}, {Boggs}, {Chiang},
  {Christensen}, {Craig}, {Giommi}  \& et al.}{{Furniss}
  et~al.}{2015}]{Furniss2015}
{Furniss} A.,  {Noda} K.,  {Boggs} S.,  {Chiang} J.,  {Christensen} F.,
  {Craig} W.,  {Giommi} P.,   et al. 2015, \mn@doi [\apj]
  {10.1088/0004-637X/812/1/65}, \href
  {https://ui.adsabs.harvard.edu/abs/2015ApJ...812...65F} {812, 65}

\bibitem[\protect\citeauthoryear{{Gallo}, {Blue}, {Grupe}, {Komossa}  \&
  {Wilkins}}{{Gallo} et~al.}{2018}]{Gallo18}
{Gallo} L.~C.,  {Blue} D.~M.,  {Grupe} D.,  {Komossa} S.,   {Wilkins} D.~R.,
  2018, \mn@doi [\mnras] {10.1093/mnras/sty1134}, \href
  {https://ui.adsabs.harvard.edu/abs/2018MNRAS.478.2557G} {478, 2557}

\bibitem[\protect\citeauthoryear{{Gao}, {Fedynitch}, {Winter}  \& {Pohl}}{{Gao}
  et~al.}{2019}]{2019NatAs...3...88G}
{Gao} S.,  {Fedynitch} A.,  {Winter} W.,   {Pohl} M.,  2019, \mn@doi [Nature
  Astronomy] {10.1038/s41550-018-0610-1}, \href
  {https://ui.adsabs.harvard.edu/abs/2019NatAs...3...88G} {3, 88}

\bibitem[\protect\citeauthoryear{{Gasparyan}, {Sahakyan}, {Baghmanyan}  \&
  {Zargaryan}}{{Gasparyan} et~al.}{2018}]{Gasp2018}
{Gasparyan} S.,  {Sahakyan} N.,  {Baghmanyan} V.,   {Zargaryan} D.,  2018,
  \mn@doi [\apj] {10.3847/1538-4357/aad234}, \href
  {https://ui.adsabs.harvard.edu/abs/2018ApJ...863..114G} {863, 114}

\bibitem[\protect\citeauthoryear{{Gasparyan}, {B{\'e}gu{\'e}}  \&
  {Sahakyan}}{{Gasparyan} et~al.}{2022}]{2022MNRAS.509.2102G}
{Gasparyan} S.,  {B{\'e}gu{\'e}} D.,   {Sahakyan} N.,  2022, \mn@doi [\mnras]
  {10.1093/mnras/stab2688}, \href
  {https://ui.adsabs.harvard.edu/abs/2022MNRAS.509.2102G} {509, 2102}

\bibitem[\protect\citeauthoryear{{Ghisellini} et~al.,}{{Ghisellini}
  et~al.}{2019}]{Ghisellini19}
{Ghisellini} G.,  et~al., 2019, \mn@doi [\aap] {10.1051/0004-6361/201935750},
  \href {https://ui.adsabs.harvard.edu/abs/2019A&A...627A..72G} {627, A72}

\bibitem[\protect\citeauthoryear{{Giann{\'\i}}, {de Rosa}, {Bassani},
  {Bazzano}, {Dean}  \& {Ubertini}}{{Giann{\'\i}} et~al.}{2011}]{Gianni2011}
{Giann{\'\i}} S.,  {de Rosa} A.,  {Bassani} L.,  {Bazzano} A.,  {Dean} T.,
  {Ubertini} P.,  2011, \mn@doi [\mnras] {10.1111/j.1365-2966.2010.17725.x},
  \href {https://ui.adsabs.harvard.edu/abs/2011MNRAS.411.2137G} {411, 2137}

\bibitem[\protect\citeauthoryear{{Giommi} \& {Padovani}}{{Giommi} \&
  {Padovani}}{2021}]{GiommiPadovani2021}
{Giommi} P.,  {Padovani} P.,  2021, arXiv e-prints, \href
  {https://ui.adsabs.harvard.edu/abs/2021arXiv211206232G} {p. arXiv:2112.06232}

\bibitem[\protect\citeauthoryear{{Giommi}, {Menna}  \& {Padovani}}{{Giommi}
  et~al.}{1999}]{Sedentary}
{Giommi} P.,  {Menna} M.~T.,   {Padovani} P.,  1999, \mn@doi [\mnras]
  {10.1046/j.1365-8711.1999.02942.x}, \href
  {https://ui.adsabs.harvard.edu/abs/1999MNRAS.310..465G} {310, 465}

\bibitem[\protect\citeauthoryear{{Giommi}, {Capalbi}, {Fiocchi}, {Memola},
  {Perri}, {Piranomonte}, {Rebecchi}  \& {Massaro}}{{Giommi}
  et~al.}{2002}]{Giommi2002babs}
{Giommi} P.,  {Capalbi} M.,  {Fiocchi} M.,  {Memola} E.,  {Perri} M.,
  {Piranomonte} S.,  {Rebecchi} S.,   {Massaro} E.,  2002, in {Giommi} P.,
  {Massaro} E.,   {Palumbo} G.,  eds, Blazar Astrophysics with BeppoSAX and
  Other Observatories. p.~63 (\mn@eprint {arXiv} {astro-ph/0209596})

\bibitem[\protect\citeauthoryear{{Giommi}, {Padovani}, {Polenta}, {Turriziani},
  {D'Elia}  \& {Piranomonte}}{{Giommi} et~al.}{2012}]{Giommi2012}
{Giommi} P.,  {Padovani} P.,  {Polenta} G.,  {Turriziani} S.,  {D'Elia} V.,
  {Piranomonte} S.,  2012, \mn@doi [\mnras] {10.1111/j.1365-2966.2011.20044.x},
  \href {https://ui.adsabs.harvard.edu/abs/2012MNRAS.420.2899G} {420, 2899}

\bibitem[\protect\citeauthoryear{{Giommi} et~al.,}{{Giommi}
  et~al.}{2019}]{Giommi2019}
{Giommi} P.,  et~al., 2019, \mn@doi [\aap] {10.1051/0004-6361/201935646}, \href
  {https://ui.adsabs.harvard.edu/abs/2019A&A...631A.116G} {631, A116}

\bibitem[\protect\citeauthoryear{{Giommi} et~al.,}{{Giommi}
  et~al.}{2020a}]{GiommiOU}
{Giommi} P.,  et~al., 2020a, in S. F.,  ed., Space Capacity Building in the XXI
  Century. Studies in Space Policy. Springer, pp 377--386 (\mn@eprint {arXiv}
  {1805.08505}), \mn@doi{https://doi.org/10.1007/978-3-030-21938-3}

\bibitem[\protect\citeauthoryear{{Giommi}, {Glauch}, {Padovani}, {Resconi},
  {Turcati}  \& {Chang}}{{Giommi} et~al.}{2020b}]{Giommi2020}
{Giommi} P.,  {Glauch} T.,  {Padovani} P.,  {Resconi} E.,  {Turcati} A.,
  {Chang} Y.~L.,  2020b, \mn@doi [\mnras] {10.1093/mnras/staa2082}, \href
  {https://ui.adsabs.harvard.edu/abs/2020MNRAS.497..865G} {497, 865}

\bibitem[\protect\citeauthoryear{{Giommi} et~al.,}{{Giommi}
  et~al.}{2021}]{GiommiXRTspectra}
{Giommi} P.,  et~al., 2021, \mn@doi [\mnras] {10.1093/mnras/stab2425}, \href
  {https://ui.adsabs.harvard.edu/abs/2021MNRAS.507.5690G} {507, 5690}

\bibitem[\protect\citeauthoryear{{HI4PI Collaboration} et~al.,}{{HI4PI
  Collaboration} et~al.}{2016}]{HI4PI}
{HI4PI Collaboration} et~al., 2016, \mn@doi [\aap]
  {10.1051/0004-6361/201629178}, \href
  {https://ui.adsabs.harvard.edu/abs/2016A%26A...594A.116H} {594, A116}

\bibitem[\protect\citeauthoryear{{Harrison} et~al.,}{{Harrison}
  et~al.}{2013}]{Harr13}
{Harrison} F.~A.,  et~al., 2013, \mn@doi [\apj] {10.1088/0004-637X/770/2/103},
  \href {http://adsabs.harvard.edu/abs/2013ApJ...770..103H} {770, 103}

\bibitem[\protect\citeauthoryear{{Hayashida} et~al.,}{{Hayashida}
  et~al.}{2015}]{Hayashida2015}
{Hayashida} M.,  et~al., 2015, \mn@doi [\apj] {10.1088/0004-637X/807/1/79},
  \href {https://ui.adsabs.harvard.edu/abs/2015ApJ...807...79H} {807, 79}

\bibitem[\protect\citeauthoryear{{IceCube Collaboration} et~al.,}{{IceCube
  Collaboration} et~al.}{2018a}]{neutrino}
{IceCube Collaboration} et~al., 2018a, \mn@doi [Science]
  {10.1126/science.aat2890}, \href
  {http://adsabs.harvard.edu/abs/2018Sci...361..147I} {361, 147}

\bibitem[\protect\citeauthoryear{{IceCube Collaboration} et~al.,}{{IceCube
  Collaboration} et~al.}{2018b}]{IceCube}
{IceCube Collaboration} et~al., 2018b, \mn@doi [Science]
  {10.1126/science.aat2890}, \href
  {https://ui.adsabs.harvard.edu/abs/2018Sci...361..147I} {361, 147}

\bibitem[\protect\citeauthoryear{{Kapanadze} et~al.,}{{Kapanadze}
  et~al.}{2016}]{Kapanadze2016}
{Kapanadze} B.,  et~al., 2016, \mn@doi [\apj] {10.3847/0004-637X/831/1/102},
  \href {https://ui.adsabs.harvard.edu/abs/2016ApJ...831..102K} {831, 102}

\bibitem[\protect\citeauthoryear{{Keivani} et~al.,}{{Keivani}
  et~al.}{2018}]{2018ApJ...864...84K}
{Keivani} A.,  et~al., 2018, \mn@doi [\apj] {10.3847/1538-4357/aad59a}, \href
  {http://adsabs.harvard.edu/abs/2018ApJ...864...84K} {864, 84}

\bibitem[\protect\citeauthoryear{{Komossa}, {Grupe}, {Parker}, {Valtonen},
  {G{\'o}mez}, {Gopakumar}  \& {Dey}}{{Komossa} et~al.}{2020}]{Komossa2020}
{Komossa} S.,  {Grupe} D.,  {Parker} M.~L.,  {Valtonen} M.~J.,  {G{\'o}mez}
  J.~L.,  {Gopakumar} A.,   {Dey} L.,  2020, \mn@doi [\mnras]
  {10.1093/mnrasl/slaa125}, \href
  {https://ui.adsabs.harvard.edu/abs/2020MNRAS.498L..35K} {498, L35}

\bibitem[\protect\citeauthoryear{{Kushwaha}, {Sarkar}, {Gupta}, {Tripathi}  \&
  {Wiita}}{{Kushwaha} et~al.}{2020}]{Kushwaha2020}
{Kushwaha} P.,  {Sarkar} A.,  {Gupta} A.~C.,  {Tripathi} A.,   {Wiita} P.~J.,
  2020, \mn@doi [\mnras] {10.1093/mnras/staa2899}, \href
  {https://ui.adsabs.harvard.edu/abs/2020MNRAS.499..653K} {499, 653}

\bibitem[\protect\citeauthoryear{{Kynoch} et~al.,}{{Kynoch}
  et~al.}{2018}]{Kynoch2018}
{Kynoch} D.,  et~al., 2018, \mn@doi [\mnras] {10.1093/mnras/stx3161}, \href
  {https://ui.adsabs.harvard.edu/abs/2018MNRAS.475..404K} {475, 404}

\bibitem[\protect\citeauthoryear{{Langejahn} et~al.,}{{Langejahn}
  et~al.}{2020}]{Lange2020}
{Langejahn} M.,  et~al., 2020, \mn@doi [\aap] {10.1051/0004-6361/202037469},
  \href {https://ui.adsabs.harvard.edu/abs/2020A&A...637A..55L} {637, A55}

\bibitem[\protect\citeauthoryear{{Lawrence} \& {Papadakis}}{{Lawrence} \&
  {Papadakis}}{1993}]{Lawrence1993}
{Lawrence} A.,  {Papadakis} I.,  1993, \mn@doi [\apjl] {10.1086/187002}, \href
  {https://ui.adsabs.harvard.edu/abs/1993ApJ...414L..85L} {414, L85}

\bibitem[\protect\citeauthoryear{{MAGIC Collaboration} et~al.,}{{MAGIC
  Collaboration} et~al.}{2008}]{MAGIC2008}
{MAGIC Collaboration} et~al., 2008, \mn@doi [Science]
  {10.1126/science.1157087}, \href
  {https://ui.adsabs.harvard.edu/abs/2008Sci...320.1752M} {320, 1752}

\bibitem[\protect\citeauthoryear{{MAGIC Collaboration} et~al.,}{{MAGIC
  Collaboration} et~al.}{2020}]{Swift2020}
{MAGIC Collaboration} et~al., 2020, \mn@doi [\aap]
  {10.1051/0004-6361/201834603}, \href
  {https://ui.adsabs.harvard.edu/abs/2020A&A...637A..86M} {637, A86}

\bibitem[\protect\citeauthoryear{{Maccacaro}, {Gioia}, {Maccagni}  \&
  {Stocke}}{{Maccacaro} et~al.}{1984}]{Maccacaro1984}
{Maccacaro} T.,  {Gioia} I.~M.,  {Maccagni} D.,   {Stocke} J.~T.,  1984,
  \mn@doi [\apjl] {10.1086/184345}, \href
  {https://ui.adsabs.harvard.edu/abs/1984ApJ...284L..23M} {284, L23}

\bibitem[\protect\citeauthoryear{{Madejski} et~al.,}{{Madejski}
  et~al.}{2016}]{Madejski2016}
{Madejski} G.~M.,  et~al., 2016, \mn@doi [\apj] {10.3847/0004-637X/831/2/142},
  \href {https://ui.adsabs.harvard.edu/abs/2016ApJ...831..142M} {831, 142}

\bibitem[\protect\citeauthoryear{{Madsen} et~al.,}{{Madsen}
  et~al.}{2015a}]{Madsen2015}
{Madsen} K.~K.,  et~al., 2015a, \mn@doi [\apjs] {10.1088/0067-0049/220/1/8},
  \href {https://ui.adsabs.harvard.edu/abs/2015ApJS..220....8M} {220, 8}

\bibitem[\protect\citeauthoryear{{Madsen} et~al.,}{{Madsen}
  et~al.}{2015b}]{Madsen2015A}
{Madsen} K.~K.,  et~al., 2015b, \mn@doi [\apj] {10.1088/0004-637X/812/1/14},
  \href {https://ui.adsabs.harvard.edu/abs/2015ApJ...812...14M} {812, 14}

\bibitem[\protect\citeauthoryear{{Madsen}, {Beardmore}, {Forster}, {Guainazzi},
  {Marshall}, {Miller}, {Page}  \& {Stuhlinger}}{{Madsen}
  et~al.}{2017}]{Madsen2017}
{Madsen} K.~K.,  {Beardmore} A.~P.,  {Forster} K.,  {Guainazzi} M.,  {Marshall}
  H.~L.,  {Miller} E.~D.,  {Page} K.~L.,   {Stuhlinger} M.,  2017, \mn@doi
  [\aj] {10.3847/1538-3881/153/1/2}, \href
  {https://ui.adsabs.harvard.edu/abs/2017AJ....153....2M} {153, 2}

\bibitem[\protect\citeauthoryear{{Mannheim}}{{Mannheim}}{1993}]{Mannheim1993}
{Mannheim} K.,  1993, \aap, \href
  {https://ui.adsabs.harvard.edu/abs/1993A&A...269...67M} {269, 67}

\bibitem[\protect\citeauthoryear{{Maraschi}, {Ghisellini}  \&
  {Celotti}}{{Maraschi} et~al.}{1992}]{Maraschi1992}
{Maraschi} L.,  {Ghisellini} G.,   {Celotti} A.,  1992, \mn@doi [\apjl]
  {10.1086/186531}, \href
  {https://ui.adsabs.harvard.edu/abs/1992ApJ...397L...5M} {397, L5}

\bibitem[\protect\citeauthoryear{{Marcha}, {Browne}, {Impey}  \&
  {Smith}}{{Marcha} et~al.}{1996}]{Marcha1996}
{Marcha} M.~J.~M.,  {Browne} I.~W.~A.,  {Impey} C.~D.,   {Smith} P.~S.,  1996,
  \mn@doi [\mnras] {10.1093/mnras/281.2.425}, \href
  {https://ui.adsabs.harvard.edu/abs/1996MNRAS.281..425M} {281, 425}

\bibitem[\protect\citeauthoryear{{Marcotulli} et~al.,}{{Marcotulli}
  et~al.}{2017}]{Marcotulli17}
{Marcotulli} L.,  et~al., 2017, \mn@doi [\apj] {10.3847/1538-4357/aa6a17},
  \href {https://ui.adsabs.harvard.edu/abs/2017ApJ...839...96M} {839, 96}

\bibitem[\protect\citeauthoryear{{Marziani}, {Sulentic}, {Dultzin-Hacyan},
  {Calvani}  \& {Moles}}{{Marziani} et~al.}{1996}]{Marziani1996}
{Marziani} P.,  {Sulentic} J.~W.,  {Dultzin-Hacyan} D.,  {Calvani} M.,
  {Moles} M.,  1996, \mn@doi [\apjs] {10.1086/192291}, \href
  {https://ui.adsabs.harvard.edu/abs/1996ApJS..104...37M} {104, 37}

\bibitem[\protect\citeauthoryear{{Massaro}, {Perri}, {Giommi}  \&
  {Nesci}}{{Massaro} et~al.}{2004a}]{Massaro2004}
{Massaro} E.,  {Perri} M.,  {Giommi} P.,   {Nesci} R.,  2004a, \mn@doi [\aap]
  {10.1051/0004-6361:20031558}, \href
  {https://ui.adsabs.harvard.edu/abs/2004A&A...413..489M} {413, 489}

\bibitem[\protect\citeauthoryear{{Massaro}, {Perri}, {Giommi}  \&
  {Nesci}}{{Massaro} et~al.}{2004b}]{2004A&A...413..489M}
{Massaro} E.,  {Perri} M.,  {Giommi} P.,   {Nesci} R.,  2004b, \mn@doi [\aap]
  {10.1051/0004-6361:20031558}, \href
  {https://ui.adsabs.harvard.edu/abs/2004A&A...413..489M} {413, 489}

\bibitem[\protect\citeauthoryear{{Massaro}, {Perri}, {Giommi}, {Nesci}  \&
  {Verrecchia}}{{Massaro} et~al.}{2004c}]{Massaro2004a}
{Massaro} E.,  {Perri} M.,  {Giommi} P.,  {Nesci} R.,   {Verrecchia} F.,
  2004c, \mn@doi [\aap] {10.1051/0004-6361:20047148}, \href
  {https://ui.adsabs.harvard.edu/abs/2004A&A...422..103M} {422, 103}

\bibitem[\protect\citeauthoryear{{Massaro}, {Tramacere}, {Cavaliere}, {Perri}
  \& {Giommi}}{{Massaro} et~al.}{2008}]{massarino2008}
{Massaro} F.,  {Tramacere} A.,  {Cavaliere} A.,  {Perri} M.,   {Giommi} P.,
  2008, \mn@doi [\aap] {10.1051/0004-6361:20078639}, \href
  {https://ui.adsabs.harvard.edu/abs/2008A&A...478..395M} {478, 395}

\bibitem[\protect\citeauthoryear{{Massaro}, {Maselli}, {Leto}, {Marchegiani},
  {Perri}, {Giommi}  \& {Piranomonte}}{{Massaro} et~al.}{2015}]{Massaro2015}
{Massaro} E.,  {Maselli} A.,  {Leto} C.,  {Marchegiani} P.,  {Perri} M.,
  {Giommi} P.,   {Piranomonte} S.,  2015, \mn@doi [\apss]
  {10.1007/s10509-015-2254-2}, \href
  {https://ui.adsabs.harvard.edu/abs/2015Ap&SS.357...75M} {357, 75}

\bibitem[\protect\citeauthoryear{{Massaro}, {Maselli}, {Leto}, {Marchegiani},
  {Perri}, {Giommi}  \& {Piranomonte}}{{Massaro} et~al.}{2016}]{Massaro2016Cat}
{Massaro} E.,  {Maselli} A.,  {Leto} C.,  {Marchegiani} P.,  {Perri} M.,
  {Giommi} P.,   {Piranomonte} S.,  2016, VizieR Online Data Catalog, \href
  {https://ui.adsabs.harvard.edu/abs/2016yCat.7274....0M} {p. VII/274}

\bibitem[\protect\citeauthoryear{{Mastichiadis} \& {Kirk}}{{Mastichiadis} \&
  {Kirk}}{2002}]{Mastichiadis2002}
{Mastichiadis} A.,  {Kirk} J.~G.,  2002, \mn@doi [\pasa] {10.1071/AS01108},
  \href {https://ui.adsabs.harvard.edu/abs/2002PASA...19..138M} {19, 138}

\bibitem[\protect\citeauthoryear{{Meyer}, {Iyer}, {Reddy}, {Georganopoulos},
  {Breiding}  \& {Keenan}}{{Meyer} et~al.}{2019}]{Meyer2019}
{Meyer} E.~T.,  {Iyer} A.~R.,  {Reddy} K.,  {Georganopoulos} M.,  {Breiding}
  P.,   {Keenan} M.,  2019, \mn@doi [\apjl] {10.3847/2041-8213/ab3db3}, \href
  {https://ui.adsabs.harvard.edu/abs/2019ApJ...883L...2M} {883, L2}

\bibitem[\protect\citeauthoryear{{Middei}, {Vagnetti}, {Bianchi}, {La Franca},
  {Paolillo}  \& {Ursini}}{{Middei} et~al.}{2017}]{Midd17}
{Middei} R.,  {Vagnetti} F.,  {Bianchi} S.,  {La Franca} F.,  {Paolillo} M.,
  {Ursini} F.,  2017, \mn@doi [\aap] {10.1051/0004-6361/201629940}, \href
  {http://adsabs.harvard.edu/abs/2017A%26A...599A..82M} {599, A82}

\bibitem[\protect\citeauthoryear{{Morrison} \& {McCammon}}{{Morrison} \&
  {McCammon}}{1983}]{Morrison1983}
{Morrison} R.,  {McCammon} D.,  1983, \mn@doi [\apj] {10.1086/161102}, \href
  {https://ui.adsabs.harvard.edu/abs/1983ApJ...270..119M} {270, 119}

\bibitem[\protect\citeauthoryear{{M{\"u}cke}, {Protheroe}, {Engel}, {Rachen}
  \& {Stanev}}{{M{\"u}cke} et~al.}{2003}]{Mucke2003}
{M{\"u}cke} A.,  {Protheroe} R.~J.,  {Engel} R.,  {Rachen} J.~P.,   {Stanev}
  T.,  2003, \mn@doi [Astroparticle Physics] {10.1016/S0927-6505(02)00185-8},
  \href {https://ui.adsabs.harvard.edu/abs/2003APh....18..593M} {18, 593}

\bibitem[\protect\citeauthoryear{{Mundo} et~al.,}{{Mundo}
  et~al.}{2020}]{Mundo2020}
{Mundo} S.~A.,  et~al., 2020, \mn@doi [\mnras] {10.1093/mnras/staa1744}, \href
  {https://ui.adsabs.harvard.edu/abs/2020MNRAS.496.2922M} {496, 2922}

\bibitem[\protect\citeauthoryear{{Murase}, {Oikonomou}  \&
  {Petropoulou}}{{Murase} et~al.}{2018}]{2018ApJ...865..124M}
{Murase} K.,  {Oikonomou} F.,   {Petropoulou} M.,  2018, \mn@doi [\apj]
  {10.3847/1538-4357/aada00}, \href
  {http://adsabs.harvard.edu/abs/2018ApJ...865..124M} {865, 124}

\bibitem[\protect\citeauthoryear{{Netzer}}{{Netzer}}{2015}]{Netzer2015}
{Netzer} H.,  2015, \mn@doi [\araa] {10.1146/annurev-astro-082214-122302},
  \href {https://ui.adsabs.harvard.edu/abs/2015ARA&A..53..365N} {53, 365}

\bibitem[\protect\citeauthoryear{Padovani \& Giommi}{Padovani \&
  Giommi}{1995}]{Padovani1995}
Padovani P.,  Giommi P.,  1995, ApJ, 444, 567

\bibitem[\protect\citeauthoryear{{Padovani} et~al.,}{{Padovani}
  et~al.}{2017}]{Padovani17}
{Padovani} P.,  et~al., 2017, \mn@doi [\aapr] {10.1007/s00159-017-0102-9},
  \href {https://ui.adsabs.harvard.edu/abs/2017A&ARv..25....2P} {25, 2}

\bibitem[\protect\citeauthoryear{{Padovani}, {Giommi}, {Resconi}, {Glauch},
  {Arsioli}, {Sahakyan}  \& {Huber}}{{Padovani} et~al.}{2018}]{Dissecting}
{Padovani} P.,  {Giommi} P.,  {Resconi} E.,  {Glauch} T.,  {Arsioli} B.,
  {Sahakyan} N.,   {Huber} M.,  2018, \mn@doi [\mnras] {10.1093/mnras/sty1852},
  \href {http://adsabs.harvard.edu/abs/2018MNRAS.480..192P} {480, 192}

\bibitem[\protect\citeauthoryear{{Padovani}, {Oikonomou}, {Petropoulou},
  {Giommi}  \& {Resconi}}{{Padovani} et~al.}{2019}]{Padovani2019}
{Padovani} P.,  {Oikonomou} F.,  {Petropoulou} M.,  {Giommi} P.,   {Resconi}
  E.,  2019, \mn@doi [\mnras] {10.1093/mnrasl/slz011}, \href
  {https://ui.adsabs.harvard.edu/abs/2019MNRAS.484L.104P} {484, L104}

\bibitem[\protect\citeauthoryear{{Paiano}, {Falomo}, {Treves}  \&
  {Scarpa}}{{Paiano} et~al.}{2018}]{Paiano2018}
{Paiano} S.,  {Falomo} R.,  {Treves} A.,   {Scarpa} R.,  2018, \mn@doi [\apjl]
  {10.3847/2041-8213/aaad5e}, \href
  {https://ui.adsabs.harvard.edu/abs/2018ApJ...854L..32P} {854, L32}

\bibitem[\protect\citeauthoryear{{Pal}, {Kushwaha}, {Dewangan}  \&
  {Pawar}}{{Pal} et~al.}{2020}]{Pal2020}
{Pal} M.,  {Kushwaha} P.,  {Dewangan} G.~C.,   {Pawar} P.~K.,  2020, \mn@doi
  [\apj] {10.3847/1538-4357/ab65ee}, \href
  {https://ui.adsabs.harvard.edu/abs/2020ApJ...890...47P} {890, 47}

\bibitem[\protect\citeauthoryear{{Paliya}, {Sahayanathan}, {Parker}, {Fabian},
  {Stalin}, {Anjum}  \& {Pandey}}{{Paliya} et~al.}{2014}]{Paliya2014}
{Paliya} V.~S.,  {Sahayanathan} S.,  {Parker} M.~L.,  {Fabian} A.~C.,  {Stalin}
  C.~S.,  {Anjum} A.,   {Pandey} S.~B.,  2014, \mn@doi [\apj]
  {10.1088/0004-637X/789/2/143}, \href
  {https://ui.adsabs.harvard.edu/abs/2014ApJ...789..143P} {789, 143}

\bibitem[\protect\citeauthoryear{{Paliya}, {B{\"o}ttcher}, {Diltz}, {Stalin},
  {Sahayanathan}  \& {Ravikumar}}{{Paliya} et~al.}{2015}]{Paliya2015}
{Paliya} V.~S.,  {B{\"o}ttcher} M.,  {Diltz} C.,  {Stalin} C.~S.,
  {Sahayanathan} S.,   {Ravikumar} C.~D.,  2015, \mn@doi [\apj]
  {10.1088/0004-637X/811/2/143}, \href
  {https://ui.adsabs.harvard.edu/abs/2015ApJ...811..143P} {811, 143}

\bibitem[\protect\citeauthoryear{{Paolillo} et~al.,}{{Paolillo}
  et~al.}{2017}]{Paolillo2017}
{Paolillo} M.,  et~al., 2017, \mn@doi [\mnras] {10.1093/mnras/stx1761}, \href
  {https://ui.adsabs.harvard.edu/abs/2017MNRAS.471.4398P} {471, 4398}

\bibitem[\protect\citeauthoryear{{Pearson}, {Unwin}, {Cohen}, {Linfield},
  {Readhead}, {Seielstad}, {Simon}  \& {Walker}}{{Pearson}
  et~al.}{1981}]{Pearson1981}
{Pearson} T.~J.,  {Unwin} S.~C.,  {Cohen} M.~H.,  {Linfield} R.~P.,  {Readhead}
  A.~C.~S.,  {Seielstad} G.~A.,  {Simon} R.~S.,   {Walker} R.~C.,  1981,
  \mn@doi [\nat] {10.1038/290365a0}, \href
  {https://ui.adsabs.harvard.edu/abs/1981Natur.290..365P} {290, 365}

\bibitem[\protect\citeauthoryear{{Pian} et~al.,}{{Pian}
  et~al.}{1998}]{pian1998}
{Pian} E.,  et~al., 1998, \mn@doi [\apjl] {10.1086/311083}, \href
  {https://ui.adsabs.harvard.edu/abs/1998ApJ...492L..17P} {492, L17}

\bibitem[\protect\citeauthoryear{{Ponti}, {Papadakis}, {Bianchi}, {Guainazzi},
  {Matt}, {Uttley}  \& {Bonilla}}{{Ponti} et~al.}{2012}]{Ponti2012}
{Ponti} G.,  {Papadakis} I.,  {Bianchi} S.,  {Guainazzi} M.,  {Matt} G.,
  {Uttley} P.,   {Bonilla} N.~F.,  2012, \mn@doi [\aap]
  {10.1051/0004-6361/201118326}, \href
  {https://ui.adsabs.harvard.edu/abs/2012A&A...542A..83P} {542, A83}

\bibitem[\protect\citeauthoryear{{Quinn} et~al.,}{{Quinn}
  et~al.}{1999}]{Quinn1999}
{Quinn} J.,  et~al., 1999, \mn@doi [\apj] {10.1086/307329}, \href
  {https://ui.adsabs.harvard.edu/abs/1999ApJ...518..693Q} {518, 693}

\bibitem[\protect\citeauthoryear{{Rani}, {Stalin}  \& {Rakshit}}{{Rani}
  et~al.}{2017}]{Rani2017}
{Rani} P.,  {Stalin} C.~S.,   {Rakshit} S.,  2017, \mn@doi [\mnras]
  {10.1093/mnras/stw3228}, \href
  {https://ui.adsabs.harvard.edu/abs/2017MNRAS.466.3309R} {466, 3309}

\bibitem[\protect\citeauthoryear{Rector, Stocke, Perlman, Morris  \&
  Gioia}{Rector et~al.}{2000a}]{Rec00}
Rector T.~A.,  Stocke J.~T.,  Perlman E.~S.,  Morris S.~L.,   Gioia I.~M.,
  2000a, AJ, 120, 1626

\bibitem[\protect\citeauthoryear{{Rector}, {Stocke}, {Perlman}, {Morris}  \&
  {Gioia}}{{Rector} et~al.}{2000b}]{Rector2000}
{Rector} T.~A.,  {Stocke} J.~T.,  {Perlman} E.~S.,  {Morris} S.~L.,   {Gioia}
  I.~M.,  2000b, \mn@doi [\aj] {10.1086/301587}, \href
  {https://ui.adsabs.harvard.edu/abs/2000AJ....120.1626R} {120, 1626}

\bibitem[\protect\citeauthoryear{{Righi}, {Tavecchio}  \& {Pacciani}}{{Righi}
  et~al.}{2019}]{2019MNRAS.484.2067R}
{Righi} C.,  {Tavecchio} F.,   {Pacciani} L.,  2019, \mn@doi [\mnras]
  {10.1093/mnras/sty3072}, \href
  {https://ui.adsabs.harvard.edu/abs/2019MNRAS.484.2067R} {484, 2067}

\bibitem[\protect\citeauthoryear{{Sahakyan}}{{Sahakyan}}{2018}]{2018ApJ...866..109S}
{Sahakyan} N.,  2018, \mn@doi [\apj] {10.3847/1538-4357/aadade}, \href
  {http://adsabs.harvard.edu/abs/2018ApJ...866..109S} {866, 109}

\bibitem[\protect\citeauthoryear{{Sahakyan}}{{Sahakyan}}{2020}]{2020A&A...635A..25S}
{Sahakyan} N.,  2020, \mn@doi [\aap] {10.1051/0004-6361/201936715}, \href
  {https://ui.adsabs.harvard.edu/abs/2020A&A...635A..25S} {635, A25}

\bibitem[\protect\citeauthoryear{{Sahakyan} \& {Giommi}}{{Sahakyan} \&
  {Giommi}}{2021}]{BLLAC2021}
{Sahakyan} N.,  {Giommi} P.,  2021, arXiv e-prints, \href
  {https://ui.adsabs.harvard.edu/abs/2021arXiv210812232S} {p. arXiv:2108.12232}

\bibitem[\protect\citeauthoryear{{Sbarrato} et~al.,}{{Sbarrato}
  et~al.}{2013}]{Sbarrato13}
{Sbarrato} T.,  et~al., 2013, \mn@doi [\apj] {10.1088/0004-637X/777/2/147},
  \href {https://ui.adsabs.harvard.edu/abs/2013ApJ...777..147S} {777, 147}

\bibitem[\protect\citeauthoryear{{Sbarrato} et~al.,}{{Sbarrato}
  et~al.}{2016}]{Sbarrato16}
{Sbarrato} T.,  et~al., 2016, \mn@doi [\mnras] {10.1093/mnras/stw1730}, \href
  {https://ui.adsabs.harvard.edu/abs/2016MNRAS.462.1542S} {462, 1542}

\bibitem[\protect\citeauthoryear{{Seielstad}, {Cohen}, {Linfield}, {Moffet},
  {Romney}, {Schilizzi}  \& {Shaffer}}{{Seielstad}
  et~al.}{1979}]{Seielstad1979}
{Seielstad} G.~A.,  {Cohen} M.~H.,  {Linfield} R.~P.,  {Moffet} A.~T.,
  {Romney} J.~D.,  {Schilizzi} R.~T.,   {Shaffer} D.~B.,  1979, \mn@doi [\apj]
  {10.1086/156929}, \href
  {https://ui.adsabs.harvard.edu/abs/1979ApJ...229...53S} {229, 53}

\bibitem[\protect\citeauthoryear{{Sikora}, {Begelman}  \& {Rees}}{{Sikora}
  et~al.}{1994}]{Sikora1994}
{Sikora} M.,  {Begelman} M.~C.,   {Rees} M.~J.,  1994, \mn@doi [\apj]
  {10.1086/173633}, \href
  {https://ui.adsabs.harvard.edu/abs/1994ApJ...421..153S} {421, 153}

\bibitem[\protect\citeauthoryear{{Sikora}, {Janiak}, {Nalewajko}, {Madejski}
  \& {Moderski}}{{Sikora} et~al.}{2013}]{Sikora2013}
{Sikora} M.,  {Janiak} M.,  {Nalewajko} K.,  {Madejski} G.~M.,   {Moderski} R.,
   2013, \mn@doi [\apj] {10.1088/0004-637X/779/1/68}, \href
  {https://ui.adsabs.harvard.edu/abs/2013ApJ...779...68S} {779, 68}

\bibitem[\protect\citeauthoryear{{Stawarz}}{{Stawarz}}{2014}]{Lukasz2014}
{Stawarz} L.,  2014, in {Ishida} M.,  {Petre} R.,   {Mitsuda} K.,  eds,
  Suzaku-MAXI 2014: Expanding the Frontiers of the X-ray Universe. p.~294

\bibitem[\protect\citeauthoryear{{Strauss}, {Huchra}, {Davis}, {Yahil},
  {Fisher}  \& {Tonry}}{{Strauss} et~al.}{1992}]{Strauss1992}
{Strauss} M.~A.,  {Huchra} J.~P.,  {Davis} M.,  {Yahil} A.,  {Fisher} K.~B.,
  {Tonry} J.,  1992, \mn@doi [\apjs] {10.1086/191730}, \href
  {https://ui.adsabs.harvard.edu/abs/1992ApJS...83...29S} {83, 29}

\bibitem[\protect\citeauthoryear{{Tavecchio} et~al.,}{{Tavecchio}
  et~al.}{2000}]{Tavecchio2000}
{Tavecchio} F.,  et~al., 2000, \mn@doi [\apj] {10.1086/317136}, \href
  {https://ui.adsabs.harvard.edu/abs/2000ApJ...543..535T} {543, 535}

\bibitem[\protect\citeauthoryear{{Turriziani}, {Fraga}  \&
  {Giommi}}{{Turriziani} et~al.}{2019}]{ST2019blaz}
{Turriziani} S.,  {Fraga} B.,   {Giommi} P.,  2019, \mn@doi [\mnras]
  {10.1093/mnras/stz2253}, \href
  {https://ui.adsabs.harvard.edu/abs/2019MNRAS.489.3307T} {489, 3307}

\bibitem[\protect\citeauthoryear{{Unwin}, {Cohen}, {Biretta}, {Pearson},
  {Seielstad}, {Walker}, {Simon}  \& {Linfield}}{{Unwin}
  et~al.}{1985}]{Unwin1985}
{Unwin} S.~C.,  {Cohen} M.~H.,  {Biretta} J.~A.,  {Pearson} T.~J.,  {Seielstad}
  G.~A.,  {Walker} R.~C.,  {Simon} R.~S.,   {Linfield} R.~P.,  1985, \mn@doi
  [\apj] {10.1086/162868}, \href
  {https://ui.adsabs.harvard.edu/abs/1985ApJ...289..109U} {289, 109}

\bibitem[\protect\citeauthoryear{{Urry} \& {Padovani}}{{Urry} \&
  {Padovani}}{1995}]{Urry95}
{Urry} C.~M.,  {Padovani} P.,  1995, \mn@doi [\pasp] {10.1086/133630}, \href
  {https://ui.adsabs.harvard.edu/abs/1995PASP..107..803U} {107, 803}

\bibitem[\protect\citeauthoryear{{Uttley} \& {McHardy}}{{Uttley} \&
  {McHardy}}{2005}]{Uttley2005}
{Uttley} P.,  {McHardy} I.~M.,  2005, \mn@doi [\mnras]
  {10.1111/j.1365-2966.2005.09475.x}, \href
  {https://ui.adsabs.harvard.edu/abs/2005MNRAS.363..586U} {363, 586}

\bibitem[\protect\citeauthoryear{{Uttley}, {McHardy}  \& {Papadakis}}{{Uttley}
  et~al.}{2002}]{Uttley2002}
{Uttley} P.,  {McHardy} I.~M.,   {Papadakis} I.~E.,  2002, \mn@doi [\mnras]
  {10.1046/j.1365-8711.2002.05298.x}, \href
  {https://ui.adsabs.harvard.edu/abs/2002MNRAS.332..231U} {332, 231}

\bibitem[\protect\citeauthoryear{{Vagnetti}, {Turriziani}  \&
  {Trevese}}{{Vagnetti} et~al.}{2011}]{Vagn11}
{Vagnetti} F.,  {Turriziani} S.,   {Trevese} D.,  2011, \mn@doi [\aap]
  {10.1051/0004-6361/201118072}, \href
  {http://adsabs.harvard.edu/abs/2011A%26A...536A..84V} {536, A84}

\bibitem[\protect\citeauthoryear{{Vagnetti}, {Middei}, {Antonucci}, {Paolillo}
  \& {Serafinelli}}{{Vagnetti} et~al.}{2016}]{Vagn16}
{Vagnetti} F.,  {Middei} R.,  {Antonucci} M.,  {Paolillo} M.,   {Serafinelli}
  R.,  2016, \mn@doi [\aap] {10.1051/0004-6361/201629057}, \href
  {http://adsabs.harvard.edu/abs/2016A%26A...593A..55V} {593, A55}

\bibitem[\protect\citeauthoryear{{Vaughan}, {Edelson}, {Warwick}  \&
  {Uttley}}{{Vaughan} et~al.}{2003}]{Vaughan2003}
{Vaughan} S.,  {Edelson} R.,  {Warwick} R.~S.,   {Uttley} P.,  2003, \mn@doi
  [\mnras] {10.1046/j.1365-2966.2003.07042.x}, \href
  {https://ui.adsabs.harvard.edu/abs/2003MNRAS.345.1271V} {345, 1271}

\bibitem[\protect\citeauthoryear{{Weekes} et~al.,}{{Weekes}
  et~al.}{1996}]{Weekes1996}
{Weekes} T.~C.,  et~al., 1996, \aaps, \href
  {https://ui.adsabs.harvard.edu/abs/1996A&AS..120C.603W} {120, 603}

\bibitem[\protect\citeauthoryear{{Weisskopf} et~al.,}{{Weisskopf}
  et~al.}{2016}]{Weisskopf2016}
{Weisskopf} M.~C.,  et~al., 2016, in {den Herder} J.-W.~A.,  {Takahashi} T.,
  {Bautz} M.,  eds,  Society of Photo-Optical Instrumentation Engineers (SPIE)
  Conference Series Vol. 9905, Space Telescopes and Instrumentation 2016:
  Ultraviolet to Gamma Ray. p. 990517, \mn@doi{10.1117/12.2235240}

\bibitem[\protect\citeauthoryear{{Whipple Collaboration: A.~C. Breslin}
  et~al.,}{{Whipple Collaboration: A.~C. Breslin} et~al.}{1999}]{whipple1999}
{Whipple Collaboration: A.~C. Breslin} et~al., 1999, arXiv e-prints, \href
  {https://ui.adsabs.harvard.edu/abs/1999astro.ph..6150W} {pp
  astro--ph/9906150}

\bibitem[\protect\citeauthoryear{Wolter \& Celotti}{Wolter \&
  Celotti}{2001}]{Wol01b}
Wolter A.,  Celotti A.,  2001, A\&A, 371, 527

\bibitem[\protect\citeauthoryear{{Yan}, {Zhang}  \& {Zhang}}{{Yan}
  et~al.}{2015}]{Da2015}
{Yan} D.,  {Zhang} L.,   {Zhang} S.-N.,  2015, \mn@doi [\mnras]
  {10.1093/mnras/stv2091}, \href
  {https://ui.adsabs.harvard.edu/abs/2015MNRAS.454.1310Y} {454, 1310}

\bibitem[\protect\citeauthoryear{{Zhang}, {Gupta}, {Gaur}, {Wiita}, {An}, {Gu},
  {Hu}  \& {Xu}}{{Zhang} et~al.}{2019}]{Zhang2019}
{Zhang} Z.,  {Gupta} A.~C.,  {Gaur} H.,  {Wiita} P.~J.,  {An} T.,  {Gu} M.,
  {Hu} D.,   {Xu} H.,  2019, \mn@doi [\apj] {10.3847/1538-4357/ab3f3a}, \href
  {https://ui.adsabs.harvard.edu/abs/2019ApJ...884..125Z} {884, 125}

\bibitem[\protect\citeauthoryear{{Zhang}, {Gupta}, {Gaur}, {Wiita}, {An}, {Lu},
  {Fan}  \& {Xu}}{{Zhang} et~al.}{2021}]{Zhang2021}
{Zhang} Z.,  {Gupta} A.~C.,  {Gaur} H.,  {Wiita} P.~J.,  {An} T.,  {Lu} Y.,
  {Fan} S.,   {Xu} H.,  2021, \mn@doi [\apj] {10.3847/1538-4357/abdd38}, \href
  {https://ui.adsabs.harvard.edu/abs/2021ApJ...909..103Z} {909, 103}

\bibitem[\protect\citeauthoryear{{de Vaucouleurs}, {de Vaucouleurs}, {Corwin},
  {Buta}, {Paturel}  \& {Fouque}}{{de Vaucouleurs}
  et~al.}{1991}]{deVaucouleurs1991}
{de Vaucouleurs} G.,  {de Vaucouleurs} A.,  {Corwin} Herold~G. J.,  {Buta}
  R.~J.,  {Paturel} G.,   {Fouque} P.,  1991, {Third Reference Catalogue of
  Bright Galaxies}

\bibitem[\protect\citeauthoryear{{in't Zand} et~al.,}{{in't Zand}
  et~al.}{2019}]{eXTP2019}
{in't Zand} J. J.~M.,  et~al., 2019, \mn@doi [Science China Physics, Mechanics,
  and Astronomy] {10.1007/s11433-017-9186-1}, \href
  {https://ui.adsabs.harvard.edu/abs/2019SCPMA..6229506I} {62, 29506}

\makeatother
\end{thebibliography}
\input{NUBLAZ_ACCEPTED.bbl} 

\clearpage

\onecolumn
\appendix
\section{Tables}

\begin{longtable}{l r r c c c c c c}
\caption{\label{listsources}\indent The 126 individual blazars resulting by the cross-matching procedure described in Sect. 2. The name of the source is taken from the corresponding catalog. The source's redshift is quoted in column 4 and column 5 accounts for their optical classification in accordance with the \textit{BZCAT} 5th edition \citep{Massaro2015,Massaro2016Cat}. We used the following code: 1= BL Lac, 2=FSRQ, 3=Blazar Uncertain type, 4= BL Lac-galaxy dominated, 5= BL Lac candidate. The SED classification relays on a visual checks and previous papers. The Obs. columns accounts for the available observations for that specific source for which alternative name is given the 8th column. Finally, the ``-`` sign is used when some information is missing.}
\setlength{\tabcolsep}{.1 pt}\\
\hline\hline
Source & Ra & Dec & z & Class & SED class & Obs & Other name \\
\hline
\endfirsthead
\caption{continued.}\\
\hline\hline
Source & Ra & Dec & z & Class & SED class & Obs & Other name \\
\hline
\endhead
\hline
\endfoot
\hline
5BZQJ0010+1058&2.62919080&10.9748622&0.089&2&LBL&1&III ZW2\\
5BZQJ0017+8135&4.28531170&81.5855933&3.387&2&LBL&3&S5 0014+81\\
5BZQJ0122+0310&20.5079583&3.1673333&4.000&2&LBL&1&3EG J0122+0310\\
5BZBJ0123-2310&20.9099580&-23.1829170&0.404&5&HBL&1&RXS J01236-2311\\
5BZQJ0126+2559&21.6783025&25.9836947&2.358&2&LBL&2&4C 25.05\\
5BZQJ0203+1134&30.9444042&11.5792806&3.639&2&LBL&1&PKS 0201+113\\
5BZGJ0214+5144&33.5747080&51.7477780&0.049&4&HBL&1&2MASX J02141794+5144520\\
5BZQJ0217+7349&34.3783883&73.8257281&2.367&2&LBL&1&HB89 0212+735\\
\end{longtable}
\section{Spectral fits}

\small{
\begin{landscape}
\setlength{\tabcolsep}{3.5pt}
\begin{longtable}{c c c c c c c c c c c c c }
		\caption{ \label{pofit} \indent Best-fit parameters derived for the 253 blazars observations presented in this paper and modelled using a power-law absorbed for the Galaxy. Columns 1 and 2 specify the name of the source and \textit{NuSTAR} sequence. Columns  3 and 4 account for the observation date and the length of the exposure. The Galactic column density  \citep[taken from][and in units of 10$^{22}$ cm$^{-2}$]{HI4PI}, the photon index, the normalisation (in units of photons keV$^{-1}$ cm$^{-2}$ s$^{-1}$) 
		and the FPMA/B inter-calibration constant are reported in columns 5, 6, 7, 8, respectively. Column 9 accounts for the ratio between the Cstat/d.o.f.~. The logarithm of the 3-10 keV (S) and 10-30 keV (H) fluxes estimated using a power-law are reported in columns 10 and 11, respectively. Finally, columns 12 and 13 refer to the maximum energy below which the spectral analysis was performed and the radius (in arcsec) of the extracting region used for the source.}
		\setlength{\tabcolsep}{.01 pt}\\
		\hline\hline
Name&ObsID&Time(MJD)&Exposure(s)&N$_{\rm H}$&$\Gamma$&Norm&Const&C/d.o.f. &log F$_{\rm S}^{\rm pl}$&log F$_{\rm H}^{\rm pl}$&E$_{\rm max}$ (keV) &Radius (arcsec)\\
		\hline
		\endfirsthead
		\caption{continued.}\\
		\hline\hline
Name&ObsID&Time(MJD)&Exposure(s)&N$_{\rm H}$&$\Gamma$&Norm&Const&C/d.o.f. &log F$_{\rm S}^{\rm pl}$&log F$_{\rm H}^{\rm pl}$&E$_{\rm max}$ (keV) &Radius (arcsec)\\
		\hline
		\endhead
		\hline
		\endfoot
		\hline
5BZQJ0010+1058&60301014002&58098.64652&93441&0.059&1.76$\pm$0.01&4.501$\pm$0.114&0.98$\pm$0.01&0.94&-10.881$\pm$0.003&-10.799$\pm$0.003&79&70\\
5BZQJ0017+8135&60001098002&57012.69861&30947&0.128&1.74$\pm$0.05&1.166$\pm$0.132&0.93$\pm$0.05&0.96&-11.460$\pm$0.009&-11.370$\pm$0.013&25&40\\
5BZQJ0017+8135&60001098004&57045.10486&36376&0.128&1.68$\pm$0.05&0.965$\pm$0.098&0.99$\pm$0.05&0.91&-11.495$\pm$0.009&-11.374$\pm$0.012&35&40\\
5BZQJ0017+8135&90201019002&57490.02500&17292&0.128&1.66$\pm$0.07&0.965$\pm$0.144&0.98$\pm$0.07&0.89&-11.479$\pm$0.015&-11.348$\pm$0.020&25&40\\
5BZQJ0122+0310&60373002002&58131.87222&30717&0.030&1.38$\pm$0.17&0.088$\pm$0.033&0.97$\pm$0.15&0.85&-12.346$\pm$0.095&-12.081$\pm$0.056&20&30\\
5BZBJ0123-2310&60101060002&57273.18125&21719&0.012&2.45$\pm$0.07&4.541$\pm$0.616&0.91$\pm$0.06&0.91&-11.382$\pm$0.010&-11.645$\pm$0.021&25&40\\
5BZQJ0126+2559&60201047002&57768.05625&41657&0.067&1.58$\pm$0.07&0.395$\pm$0.056&1.00$\pm$0.06&0.92&-11.800$\pm$0.013&-11.627$\pm$0.016&25&40\\
5BZQJ0126+2559&60367001002&58121.66736&19830&0.067&1.60$\pm$0.11&0.341$\pm$0.082&1.09$\pm$0.11&0.79&-11.884$\pm$0.017&-11.723$\pm$0.036&20&40\\
\end{longtable}
\end{landscape}

\begin{landscape}
\setlength{\tabcolsep}{3.5pt}
\begin{longtable}{c c c c c c c c c c c c c c }
		\caption{\label{logparfit}\indent Best-fit parameters derived for the 253 blazars observations modelled using a logarithmic parabola absorbed for the Galaxy. Columns 1 and 2 specify the object name and \textit{NuSTAR} observation ID. Columns from 3 and 4 account for the observation date and the length of the exposure. The Galactic column density \citep[taken from][and in units of 10$^{22}$ cm$^{-2}$]{HI4PI}, $\alpha$, $\beta$, the normalisation and the FPMA/B inter-calibration constants are reported in columns 5, 6, 7, 8, and 9, respectively. Then, column 10 accounts for the ratio between the Cstat/d.o.f.. The logarithm of the 3-10 keV (S) and 10-30 keV (H) fluxes (in units of 1/erg cm$^{-2}$ s$^{-1}$) estimated using the logarithmic-parabola are reported in columns 11 and 12, respectively. Finally, columns 13 and 14 refer to the maximum energy below which the spectral analysis was performed and the radius (in arcsec) of the extracting region used for the source.}
		\setlength{\tabcolsep}{.01 pt}\\
		\hline\hline
Name&ObsID&Time(MJD)&Exposure(s)&N$_{\rm H}$&$\alpha$&$\beta$&Norm&Const&C/d.o.f. &log F$^{\rm logpar}_{\rm S}$&log F$^{\rm logpar}_{\rm H}$&E$_{\rm max}$ (keV)& Radius (arcsec)\\
		\hline
		\endfirsthead
		\caption{continued.}\\
		\hline\hline
Name&ObsID&Time(MJD)&Exposure(s)&N$_{\rm H}$&$\alpha$&$\beta$&Norm&Const&C/d.o.f. &log F$^{\rm logpar}_{\rm S}$&log F$^{\rm logpar}_{\rm H}$&E$_{\rm max}$ (keV)& Radius (arcsec)\\
		\hline
		\endhead
		\hline
		\endfoot
		\hline
5BZQJ0010+1058&60301014002&58098.64652&93441&0.059&1.78$\pm$0.02&-0.047$_{-0.037}^{+0.037}$&2.66$\pm$0.03&0.98$\pm$0.01&0.94&-10.881$\pm$0.002&-10.796$\pm$0.004&79&70\\
5BZQJ0017+8135&60001098002&57012.69861&30947&0.128&1.73$\pm$0.11&0.039$_{-0.223}^{+0.227}$&0.70$\pm$0.03&0.93$\pm$0.05&0.96&-11.460$\pm$0.009&-11.374$\pm$0.020&25&40\\
5BZQJ0017+8135&60001098004&57045.10486&36376&0.128&1.65$\pm$0.09&0.070$_{-0.175}^{+0.179}$&0.64$\pm$0.03&0.99$\pm$0.05&0.91&-11.495$\pm$0.009&-11.379$\pm$0.016&35&40\\
5BZQJ0017+8135&90201019002&57490.02500&17292&0.128&1.52$\pm$0.14&0.338$_{-0.298}^{+0.304}$&0.67$\pm$0.04&0.98$\pm$0.07&0.89&-11.478$\pm$0.013&-11.378$\pm$0.024&25&40\\
5BZQJ0122+0310&60373002002&58131.87222&30717&0.030&1.49$\pm$0.34&-0.289$_{-0.767}^{+0.807}$&0.10$\pm$0.01&0.97$\pm$0.15&0.85&-12.304$\pm$0.027&-11.998$\pm$0.069&20&30\\
5BZBJ0123-2310&60101060002&57273.18125&21719&0.012&2.41$\pm$0.11&0.141$_{-0.312}^{+0.322}$&0.89$\pm$0.04&0.91$\pm$0.06&0.91&-11.382$\pm$0.010&-11.662$\pm$0.028&25&40\\
5BZQJ0126+2559&60201047002&57768.05625&41657&0.067&1.49$\pm$0.14&0.191$_{-0.277}^{+0.283}$&0.31$\pm$0.02&1.00$\pm$0.06&0.92&-11.800$\pm$0.011&-11.644$\pm$0.029&25&40\\
5BZQJ0126+2559&60367001002&58121.66736&19830&0.067&1.68$\pm$0.22&-0.209$_{-0.516}^{+0.532}$&0.26$\pm$0.02&1.09$\pm$0.11&0.79&-11.885$\pm$0.022&-11.698$\pm$0.044&20&40\\
\end{longtable}
\end{landscape}
}

\bsp	
\label{lastpage}
\end{document}